\documentclass{article}
\usepackage[a4paper, margin=1in]{geometry}
\usepackage{graphicx} 
\usepackage[utf8]{inputenc}
\usepackage{amsmath}
\usepackage{algorithm}
\usepackage{algpseudocode}
\usepackage{amsthm}
\usepackage{amsfonts}
\usepackage{hyperref}
\usepackage{xcolor}
\usepackage{aas_macros}

\newcommand{\real}{\operatorname{Re}}

\newcommand{\dmass}{\hat{\mu}} 
\newcommand{\denergy}{\hat{\mathcal{E}}} 
\newcommand{\diff}{\mathcal{D}}
\newcommand{\bR}{\mathbb{R}}
\newcommand{\bC}{\mathbb{C}}

\title{Efficient High-order Mass-conserving and Energy-balancing Schemes for Schr\"{o}dinger-Poisson Equations}
\author{
Manvendra Pratap Rajvanshi\textsuperscript{1}\thanks{\url{manvendra.rajvanshi@kaust.edu.sa}} and
David I. Ketcheson\textsuperscript{1}\thanks{\url{david.ketcheson@kaust.edu.sa}}\\[2pt]\small\textsuperscript{1}CEMSE Division, King Abdullah University of Science and Technology, Thuwal, Saudi Arabia 
}
\date{April, 2026}

\begin{document}

\maketitle

\begin{abstract}
    We study relaxation-based approaches for conserving mass and energy in the numerical solution of Schr\"{o}dinger-Poisson (SP) type systems. Relaxation-based methods offer a general approach that can be applied as post-time step processing to achieve conservation with any time-stepping scheme. Here we study two types of relaxation techniques applied to implicit-explicit Runge-Kutta schemes, with Fourier collocation in space.  We also study SP equations with time-varying coefficients (which appear naturally in cosmology) where energy is not conserved but satisfies a balance equation.
    We show that the fully-discrete system conserves both mass and energy (or satisfies the balance equation in case of time-varying coefficients), up to rounding errors.  The effectiveness of these methods is demonstrated via numerical examples, including a three-dimensional cosmological simulation.
\end{abstract}

\section{Introduction}

Schr\"{o}dinger-Poisson (also known as Schr\"{o}dinger-Newton) equations are used to model nonlinear phenomena in various fields, including nonlinear optics in thermo-optical media \cite{PAREDES2020132301}, semiconductor applications \cite{10.5555/89817}, and self-gravitating Bose-Einstein condensates (BECs) \cite{PhysRevD.92.124008}.  For a  self-gravitating system of large number of bosons, mean-field approximation can be used to derive Gross-Pitaevskii-Newton equation that can be further simplified to SP equations under certain assumptions \cite{PhysRevD.84.043531}. In cosmology, they can be employed as an approximation of the Vlasov-Poisson system for collisionless self-gravitating particles \cite{1993ApJ...416L..71W}, via Husimi representation \cite{husimi}.  Alternatively, since SP models self-gravitating bosons \cite{PhysRev.187.1767}, it arises directly in certain scalar field models of dark matter, like fuzzy dark matter \cite{PhysRevLett.85.1158} or dark matter from ultra-light axions \cite{MARSH20161}.

Due to their wide range of applications, there has been a lot of interest in developing computational methods for the SP equations, from both the numerical mathematics community and application domain experts.
We consider the following class of Schr\"{o}dinger-Poisson equations \cite{athanassoulis2023novel}: 
\begin{subequations}\label{eq:SP}
\begin{align}
\boldsymbol{\psi}_t - i p(t)\nabla^2 \boldsymbol{\psi} + i q(t) \boldsymbol{V} \boldsymbol{\psi} &= 0 \label{eq:SP_a} \\[6pt]
\nabla^2 \boldsymbol{V} &= |\boldsymbol{\psi}|^2 - \mu. \label{eq:SP_b}
\end{align}
\end{subequations}
Here $\psi(t,x)$ is complex-valued ($[0,T]\times \bR^d\to \bC$) and $V(t,x)$ is real-valued ($[0,T]\times \bR^d\to \bR$) with $d \in \{1,2,3\}$. The most important physical quantities in this system are the total mass and total energy, represented by\footnote{Here and throughout, we use $()^*$ to denote complex conjugation and $()^\dagger$ for Hermitian transpose.}
\begin{align}
   \mu & =  \|\boldsymbol{\psi}\|^2 \equiv  \int \boldsymbol{\psi}^*\boldsymbol{\psi} dx \label{eq:mass} \\
    \mathcal{E} & = p(t) E_k-\frac{q(t)}{2} E_v \label{eq:energy}
\end{align}
where
\begin{equation*}
    E_k = \|\nabla \boldsymbol{\psi} \|^2, \qquad E_v = \|\nabla \boldsymbol{V} \|^2.
\end{equation*}
In general, the mass is conserved while the energy satisfies a balance equation:
\begin{align}
\frac{d}{dt}\mu & = 0 \\
p \frac{dE_k}{dt} - \frac{q}{2} \frac{dE_v}{dt} & = 0.
\label{eq:energy_eqn}
\end{align}
Note that
if $p$ and $q$ are constant in time, eq. \eqref{eq:energy_eqn} reduces to the usual energy conservation law.

\subsection{Structure-preserving numerical methods}
In order to maintain the physical properties mentioned
above, it is desirable to design numerical schemes that
\begin{itemize}
    \item conserve mass;
    \item conserve energy in the case of constant $p$ and $q$;
    \item preserve the energy balance law \eqref{eq:energy_eqn} in general.
\end{itemize}
There has been significant work in this direction.

We do not review the literature on non-conservative SP schemes here, but instead refer the reader to 
the introductions of the papers by Wang et. al. \cite{wang2025point} and Athanassoulis et. al.\cite{athanassoulis2023novel} which provide a fairly thorough list of such references.

Operator-splitting schemes for SP (e.g. \cite{M2AN_2017__51_4_1245_0,Bao2003809,Cheng2022523}), where the time-evolution operator is split into a linear and a nonlinear operator, typically conserve mass but not energy.
One of the first works on schemes preserving both mass and energy is that of Ringhofer \& Soler \cite{ringhofer2000discrete}, who use summation by parts (SBP) in space and the implicit trapezoidal method (i.e. Crank-Nicolson) in time.
Ehrhardt \& Zisowsky \cite{ehrhardt2006fast} use the method of Ringhofer in 3D with spherical
symmetry and introduce an approach for non-reflecting boundary conditions.
Subsequent works have mostly followed a similar approach,
pairing a  conservative spatial discretization with an implicit time integrator that maintains conservation. This strategy typically yields methods that are second-order accurate in time while permitting higher-order accuracy in space.  For example,
Cheng et. al \cite{8466572} propose a finite-difference based scheme that is second order (Crank-Nicolson) in time and fourth order in space.  Gong et. al \cite{SAV} reformulate the original SP system using a scalar auxiliary variable (SAV) and then use Crank-Nicolson in time and Galerkin-Legendre spectral spatial discretization.
Whereas the foregoing schemes are fully (nonlinearly) implicit, 
Nemati et. al. \cite{nemati2025high} give a linearly implicit scheme using 4th-order compact finite differences in space
and operator splitting in time (with Crank-Nicolson for the stiff term), employing the alternating
direction implicit (ADI) strategy to reduce the cost of multidimensional solves.  
Wang et. al \cite{wang2025point} use finite-difference based implicit schemes that are second order in both space and time.
Athanassoulis et. al. \cite{athanassoulis2023novel} give a scheme that is not only mass and energy conserving, but also satisfies a discrete energy balance in the general case; it is a finite element scheme that uses an auxiliary variable and is only
linearly implicit and second-order in time.  
Their approach is similar to that
of earlier methods for NLS \cite{10.1093/imanum/drz067}.
Mina et. al \cite{2020A&A...641A.107M} solve an additional continuity equation and use rescaling using the solution of this additional equation to conserve mass and satisfy energy balance.

The foregoing schemes 
are limited to second-order accuracy in time, and most of them are not designed to preserve energy balance in the general case of time-varying $p$ and $q$. 
In the present work we propose and test a class of methods that combines higher order Implicit-Explicit (ImEx) Runge-Kutta (RK) methods \cite{doi:10.1137/1.9781611978209} with relaxation in time \cite{ketcheson2019relaxationrungekuttamethodsconservation,Ranocha_2020, doi:10.1137/23M1598118,ranocha2025highordermassenergyconservingmethods}. These methods conserve both mass and energy (or satisfy the proper energy balance in the non-conservative case) and are high-order in time.  They share with some other recently-proposed schemes the advantage of being implicit only in the linear term, avoiding the need to solve large nonlinear algebraic systems.

Relaxation is a technique similar to projection, in
which after each step the solution is perturbed back
onto the conservative manifold.
Early works using relaxation \cite{SANZSERNA1982199,SANZSERNA1983273} used leapfrog time integration with relaxation for KdV and Nonlinear Schr\"{o}dinger equations (NLS). The concept has since been refined and extended to different time-stepping schemes including Runge-Kutta (RK) methods \cite{ketcheson2019relaxationrungekuttamethodsconservation} and multistep schemes \cite{Ranocha_2020}. More recently
it has been extended to enforce conservation of multiple invariants, either through multiple relaxation \cite{doi:10.1137/23M1598118} or by combining it with projection \cite{ranocha2025highordermassenergyconservingmethods}. 
In contrast to the implicit time stepping methods above, relaxation requires the solution of only one or two algebraic equations at each time step.
In this work we adapt the ideas of multiple-relaxation and projection-relaxation (\cite{doi:10.1137/23M1598118} and \cite{ranocha2025highordermassenergyconservingmethods}) and combine them with ImEx schemes to obtain higher order (in time) methods that satisfy both mass conservation and energy balance for SP equations.

Although here we work with ImEx RK methods in time and pseudospectral Fourier collocation methods in space, our approach could be applied to any method as long as the spatial discretization is mass- and energy-conserving. 
Thus one can leverage the benefits of relaxation techniques for enhanced conservation properties by combining them with their choice of numerical method.

A major difficulty with implicit conservative schemes is their computational cost, since they typically require the solution of either nonlinear systems or very large linear systems at each time step. As noted by Wang et. al \cite{wang2025point}, the central challenge is to preserve the discrete conservation laws without compromising computational efficiency during time integration while ensuring computational efficiency. In the methodology proposed in this paper, the implicit updates are carried out using fast Fourier transforms (FFTs), which are computationally efficient. 

The rest of this article is organized as follows.
Our primary motivation is the application of SP to cosmological models of an expanding universe, which we briefly describe in Section \ref{subsec:cosmo_expand}.
In Section \ref{sec:discrete_conserve}, we show that spatial  discretization of SP satisfies discrete analogs of the mass conservation law and the energy balance law \eqref{eq:energy_eqn} as long as the discrete first-derivative operator is skew-Hermitian. 
We describe our time discretization, including multiple relaxation and projection relaxation, in Section \ref{sec:time-disc}; Section \ref{subsec:energy_enforce} contains an explanation of how the energy-balance law is enforced discretely. We present numerical test cases in Section \ref{sec:numerical_cases}, showing that 
the proposed methodology not only conserves mass and energy but also enhances the overall quality of the solution.
Some conclusions are given in Section \ref{sec:conclusion}.

\subsection{Schr\"{o}dinger-Poisson in an Expanding Universe}
\label{subsec:cosmo_expand}
In the context of cosmology and astrophysics, SP equations find application to multiple modeling problems. They are used as a tool for approximating Vlasov-Poisson (VP) systems for simulating collisionless dark matter; see e.g.  \cite{Kopp_2017,1993ApJ...416L..71W,PhysRevD.97.083519} for studies of the correspondence between SP and VP solutions. 

SP equations can also be used to represent dynamics of the fundamental scalar fields that might play different roles in dynamics of the universe. A number of dark matter models can be effectively treated as scalar field models, and their effective dynamics can be represented by SP equations \cite{MARSH20161}. One concrete example is that of ultra-light axions in axion cosmology \cite{MARSH20161} in which the SP equations model a classical field arising out of a very large number of extremely low mass axion particles. At a phenomenological level, SP equations are used to model a general form of dark matter called fuzzy dark matter (FDM) \cite{PhysRevLett.85.1158} or wavelike dark matter \cite{Schive_2014}. These dark matter models exhibit astrophysical scale interference patterns and can lead to halos with solitonic cores \cite{PhysRevLett.113.261302} that might help alleviate the core-cusp problem \cite{de_Blok_2009,10.1093/mnras/stv1050}. Furthermore, this quantum-wave-like model suppresses structure formation at small scales that can potentially help resolve some challenges faced by standard cold dark matter (CDM) \cite{Zhang_2019}. This has provided ample motivation for numerical studies of fuzzy dark matter using Schr\"{o}dinger-Poisson equations \cite{Zhang_2019,schive2025fuzzydarkmattersimulations}.

In terms of numerical discretization, pseudospectral collocation with some kind of operator splitting appears to be the approach most common among the astrophysical/cosmological community \cite{schive2025fuzzydarkmattersimulations}.  Most of these schemes are second order in time \cite{10.1093/mnras/stab1764,Mocz_2017,Edwards_2018,Woo_2009}, although notably Levkov et. al \cite{PhysRevLett.121.151301} and Schwabe et. al \cite{PhysRevD.102.083518} use a sixth-order operator splitting scheme. For a recent review of FDM simulations using SP, see \cite{schive2025fuzzydarkmattersimulations}.  

Here we write the equations as they appear in the cosmological setting and then we connect them to the form stated in \eqref{eq:SP}. The Schr\"{o}dinger-Poisson equation for an expanding universe with relevant physical coefficients shown explicitly takes the form
\begin{subequations}\label{eq:SP_cosmo}
\begin{align}
\frac{\partial \boldsymbol{\psi}}{\partial a} &= \frac{i\epsilon}{2\mathcal{H}a^3}\nabla_c^2 \boldsymbol{\psi} - \frac{i\beta}{\epsilon\mathcal{H}a^2} \boldsymbol{V} \boldsymbol{\psi} \label{eq:SP_cosmo_a} \\
\nabla_c^2 \boldsymbol{V} &= |\boldsymbol{\psi}|^2 - 1, \label{eq:SP_cosmo_b}
\end{align}
\end{subequations}
with the following physical meanings:
\begin{equation}
    |\boldsymbol{\psi}|^2 = (1+\delta), \quad\epsilon = \frac{\hbar}{mH_0},\quad\mathcal{H}=\frac{H}{H_0}.
\end{equation}
Here $\delta$ is density contrast, $a$ is the cosmological expansion factor, $\hbar$ is the reduced Planck constant, $m$ is mass of particle, $H$ is Hubble parameter and $H_0$ is Hubble constant (i.e. Hubble parameter at present). 
Eq. \eqref{eq:SP_cosmo} is equivalent to \eqref{eq:SP} with the following correspondences:
\begin{align*}
    t & = a &
    p & = \frac{\epsilon}{2\mathcal{H}a^3} &
    q & = \frac{\beta}{\epsilon\mathcal{H}a^2.}
\end{align*}
Because of the dynamics in expanding space, energy is not conserved; instead it satisfies an energy balance relation known as the Layzer-Irvine equation \cite{1961PhDT.........2I,1963ApJ...138..174L}. In case of \eqref{eq:SP_cosmo_a}, this takes the following form:
\begin{equation}
\frac{\epsilon}{\mathcal{H}a^3} \frac{dE_k}{dt} - \frac{\beta}{\epsilon\mathcal{H}a^2} \frac{dE_v}{dt} = 0.
\label{eq:LzIr_a}
\end{equation}

\section{Mass-Conservative and Energy-Balanced Spatial Discretization}
\label{sec:discrete_conserve}
We consider the problem in one space dimension, and
discretize the spatial domain, approximating $\psi$ by
the vector
$\psi(t)=[\psi_0(t),\psi_1(t),...\psi_{N-1}(t)]^T$ and potential by the vector $V$ \footnote{We use $\psi$ to denote the discretized counterpart of $\boldsymbol{\psi}$ and $V$ to denote the discretized counterpart of $\boldsymbol{V}$.}. The spatial derivative ($\nabla$) is approximated by a matrix $\mathcal{D}$; for now we only assume that this matrix is skew-Hermitian:
\begin{align} \label{eq:skewh}
\mathcal{D}^\dagger=-\mathcal{D}.
\end{align}
This gives us the semi-discretized system
\begin{subequations}\label{eq:main}
\begin{align}
\psi_t - i p(t)\mathcal{D}^2 \psi + i q(t) L \psi &= 0 \label{eq:SP_a_disc} ,\qquad \mathop{diag}(L)=V\\
\mathcal{D}^2 V &= \mathop{Vec}(|\psi|^2) - \mu \label{eq:SP_b_disc}
\end{align}
\end{subequations}
with the discrete analog of kinetic and potential energy:
\[
\bar{E}_k = \langle \mathcal{D} \psi, \mathcal{D} \psi \rangle, 
\qquad \bar{E}_v = \langle \mathcal{D} V, \mathcal{D} V \rangle.
\]
Here the inner product is defined as  
$\langle a,b\rangle = \sum_j a^*_jb_j$.
We define the discrete analogs of mass and energy as follows:
\begin{align*}
\dmass & = \|\psi\|^2 \\
\denergy & = p \| \diff \psi \|^2 - \frac{q}{2} \|\diff V\|^2.
\end{align*}

Let us now consider how these discrete quantities evolve in time.
By taking the inner product of \eqref{eq:SP_a_disc} with $\psi$, we get 
\begin{equation}
    \langle \psi, \psi_t \rangle 
+ i p \|\mathcal{D}\psi\|^2
+ i q \langle \psi, L \psi \rangle = 0.
\label{eq:disc_mass_conserve_a}
\end{equation}
Noting that $p$,$q$ and $\langle \psi, L \psi \rangle $ are real and taking real part of the above equation, we get discrete mass conservation
\begin{equation}
    \real\langle \psi, \psi_t \rangle = 0 \implies \langle \psi, \psi \rangle_t = 0.
    \label{eq:disc_mass_conserve_b}
\end{equation}
Meanwhile, the discrete energies satisfy a dynamics equation analogous to
\eqref{eq:energy_eqn}.
To see this, we take the inner product of \eqref{eq:SP_a_disc} with $\psi_t$:
\begin{align}
\langle \psi_t, \psi_t \rangle 
- i p \langle \psi_t, \mathcal{D}^2 \psi \rangle 
+ i q \langle \psi_t, L \psi \rangle &= 0 \\[6pt]
\|\psi_t\|^2 
- i p \langle \mathcal{D}^\dagger \psi_t, \mathcal{D} \psi \rangle 
+ i q \langle \psi_t, L \psi \rangle &= 0 
\label{eq:1}
\end{align}
Taking the imaginary part of the above equation \eqref{eq:1},  and using \eqref{eq:skewh} we obtain
\begin{align}
- i p \, \real \langle \mathcal{D}^\dagger \psi_t, \mathcal{D} \psi \rangle 
+ i q \, \real \langle \psi_t, L \psi \rangle = 0 \\
p \, \real \langle \mathcal{D} \psi_t, \mathcal{D} \psi \rangle 
+ q \, \real \langle \psi_t, L \psi \rangle = 0. \label{eq:realparts}
\end{align}

From the product rule we obtain the identity
\begin{equation}
    \langle \mathcal{D} \psi, \mathcal{D} \psi \rangle_t 
= 2 \, \real \langle \mathcal{D} \psi_t, \mathcal{D} \psi \rangle.
\label{eq:id_1}
\end{equation}
Substituting this in \eqref{eq:realparts}, we get
\begin{equation}
     \quad 
\frac{p}{2} \langle \mathcal{D} \psi, \mathcal{D} \psi \rangle_t 
+ q \, \real \langle \psi_t, L \psi \rangle = 0.
\label{eq:2}
\end{equation}
Next we take \eqref{eq:SP_b_disc}, differentiate in time, and take the inner product with $V$ to obtain (recalling that $\mu$ is constant in time on the discrete level)
\begin{equation}
\langle V, (\mathcal{D}^2 V)_t \rangle = \real \langle V, 2 \psi\, \psi_t^* \rangle.
\end{equation}
Using \eqref{eq:skewh}, we have
\begin{equation}
- \langle \mathcal{D} V, \mathcal{D} V_t \rangle 
= \langle V, \real(2 \psi\, \psi_t^*) \rangle,
\end{equation}
and using identity \eqref{eq:id_1}
\begin{align}
\Rightarrow \; -\frac{d}{dt} \frac{\|\mathcal{D} V\|^2}{2} 
&= \langle V, \real(2 \psi\, \psi_t^*) \rangle 
= 2 \real \langle \psi_t, L \psi \rangle \\[6pt]
\Rightarrow \; \real \langle \psi_t, L \psi \rangle 
&= - \frac{1}{4} \frac{d}{dt} \|\mathcal{D} V\|^2.
\label{eq:3}
\end{align}
Using equation \eqref{eq:3} in \eqref{eq:2}, we get
\begin{align}
\frac{p}{2} \langle \mathcal{D} \psi, \mathcal{D} \psi \rangle_t 
- \frac{q}{4} \frac{d}{dt} \|\mathcal{D} V\|^2 &= 0 \\[6pt]
\Rightarrow \; 
\frac{p}{2} \frac{d}{dt} \langle \mathcal{D} \psi, \mathcal{D} \psi \rangle 
- \frac{q}{4} \frac{d}{dt} \langle \mathcal{D} V, \mathcal{D} V \rangle &= 0
\end{align}
which is the conservation law:
\begin{equation}
\label{eq:dis_eng}
\boxed{\;
p \frac{d\bar{E}_k}{dt} - \frac{q}{2} \frac{d\bar{E}_v}{dt} = 0.
\;}  
\end{equation}

\section{Time Discretization}
\label{sec:time-disc}
In this section we describe the temporal discretization that we apply to \eqref{eq:SP}.  First, in Section \ref{sec:imex} we briefly review implicit-explicit Runge-Kutta (ImEx RK) methods.  Then in Sections \ref{sec:MR} and \ref{sec:PR} we review the ideas of multiple relaxation and projection-relaxation, which we use to enforce mass  and energy conservation and energy balance.  Finally, in Section
\ref{subsec:energy_enforce}, we detail our new approach to adapt these techniques to enforce the energy balance equation for the case of varying $p(t)$ and $q(t)$.

\subsection{ImEx RK methods} \label{sec:imex}
To integrate \eqref{eq:main} in time, we first divide the evolution terms into two parts, $f$ and $g$:
\begin{equation}
    \psi_t  = \underbrace {i p(t)\nabla^2 \psi}_{=g(\psi)} - \underbrace{i q(t) L \psi}_{=f(\psi)}  
\end{equation}
We then apply an
Implicit Explicit Runge Kutta (ImEx) method \cite{doi:10.1137/1.9781611978209},
integrating the nonstiff term $f$ explicitly and the
stiff term $g$ implicitly: 
\begin{align}
\psi^{(i)} &= \psi^n + \Delta t \left( \sum_{j=1}^{i-1} \tilde{a}_{ij} f(\psi^{(j)}) + \sum_{j=1}^{i} a_{ij} g(\psi^{(j)}) \right), \quad i = 1, 2, \dots, s, \\
\psi^{n+1} &= \psi^n + \Delta t \left( \sum_{j=1}^{s} \tilde{b}_{j} f(\psi^{(j)}) + \sum_{j=1}^{s} b_{j} g(\psi^{(j)}) \right).
\end{align}
In the numerical examples below, we use the 3rd order method ARK3(2)4L[2]SA and the 4th order method ARK4(3)6L[2]SA from Kennedy \& Carpenter \cite{kennedy2003additive}.

\subsection{Multiple Relaxation (MR)}
\label{sec:MR}
Here we closely follow the notation and methodology of Biswas et. al \cite{doi:10.1137/23M1598118}. We describe the modifications that are done at each time step. Consider that we are solving a differential equation of the form:
\begin{equation}
    \Dot{\psi}(t) = f(t,\psi(t)) + g(t,\psi(t))
\end{equation}
with an intent to conserve $\dmass$ and $\denergy$. Starting from a numerical solution $\psi^n\approx \psi(t_n)$, we apply an embedded ImEx scheme to calculate the update vectors $d_1^n$, $d_2^n$, from two embedded methods that share same intermediate stages but have different weights $\vec{b}^1, \vec{b}^2$:
\begin{equation}
d_k^n := \sum_{j=1}^{s} \tilde{b}_{j}^{k}\, f\!\left(t_n + \tilde{c}_j \Delta t,\; \psi^{(j)}\right) +\sum_{j=1}^{s} b_{j}^{k}\, g\!\left(t_n + c_j \Delta t,\; \psi^{(j)}\right),
\qquad \text{for } k = 1, 2.
\end{equation}
Then the final update for this time step is done via:
\begin{equation}
\psi\!\left(t_n + (1+\Gamma)\,\Delta t\right)
\approx \psi^{n+1}_{\gamma}
:= \psi^{n+1} + \Delta t \sum_{i=1}^{2} \gamma_i\, d_i^{n}.
\end{equation}
where $\Gamma = \sum_i\gamma_i$, $\psi^{n+1}:=\psi^n+(\Delta t) d^n_1$ and $\vec{\gamma}$ is chosen so as to satisfy discrete mass and energy conservation:
\begin{subequations} \label{eq:MR-system}
\begin{align}
    \dmass(\psi^{n+1}_{\gamma}) & = \dmass(\psi^n) \label{eq:MR-system_a}  \\
    \denergy(\psi^{n+1}_{\gamma}) & = \denergy(\psi^n) \label{eq:MR-system_b}
\end{align}
\end{subequations}
Note that the time is updated from $t_n$ to $t_n + (1+\Gamma)(\Delta t)$ instead of $t_n+\Delta t$. 
Thus the multiple relaxation approach requires us to solve a system of two algebraic equations at each time step.

\subsection{Projection Relaxation (PR)}
\label{sec:PR}
This method was introduced in Ranocha et. al \cite{ranocha2025highordermassenergyconservingmethods} for mass and energy conservation for NLS and its hyperbolization \cite{https://doi.org/10.1111/sapm.70129}. In contrast to the multiple-relaxation method of the previous subsection, this technique requires only solving a scalar nonlinear optimization problem even when conserving both mass and energy. The technique is similar; we first calculate an updated approximation $\psi^{n+1} 
\approx \psi(t^n+\Delta t)$ using a standard time stepping method (in our case, an ImEx Runge-Kutta method). Next, this solution is orthogonally projected onto the mass-conserving manifold; we represent the projection operator by $\pi$:
\begin{equation}
    \hat{\psi}^{\,n+1} = \pi\!\left(\psi^{\,n+1}\right)
\end{equation}
Then we solve a relaxation-type equation that enforces energy conservation while remaining on the mass-conservative manifold.  Recall that $\denergy$ denotes the (discrete) total energy; then we solve
\begin{equation}
    \denergy\!\left(
\pi\!\left(
\psi^{n} + \gamma\left(\hat{\psi}^{\,n+1} - \psi^{n}\right)
\right)
\right)
= \denergy\!\left(\psi^{n}\right) \label{eq:projRlx_opt}
\end{equation}
Finally, the updated solution is given by
\begin{equation}
    \psi_\gamma^{n+1}
= \pi\!\left(\psi^{n} + \gamma\left(\hat{\psi}^{\,n+1} - \psi^{n}\right)\right)
\approx \psi\!\left(t^{n+1}\right),
\qquad
t^{n+1} = t^{n} + \gamma\,\Delta t.
\end{equation}
The projection operator $\pi(\psi)$ is:
\begin{equation}
    \pi\!\left(\psi\right)
=
\sqrt{\frac{\dmass\!\left(\psi^{n}\right)}
           {\dmass\!\left(\psi\right)}}
\,\psi.
\end{equation}

\subsection{Enforcing Energy Balance via Relaxation}
\label{subsec:energy_enforce}
As mentioned earlier, in the case of cosmological applications of SP, the energy satisfies a balance law instead of a conservation equation. 
Both of the approaches just described can be adapted to ensure accurate energy dynamics in this setting.

In order to implement energy balance via relaxation, we need a time-discretized version of the energy-balance equation \eqref{eq:dis_eng}.
By the fundamental theorem of calculus, the energy satisfies
\begin{equation}
    \int_{t_n}^{t_n+\Delta t}\frac{d\mathcal{E}}{dt}dt = \mathcal{E}(t_n+\Delta t)-\mathcal{E}(t_n)
\end{equation}
If we consider $\mathcal{E} = \left(pE_k-\frac{1}{2}qE_v\right)$ and use the energy balance law \eqref{eq:energy_eqn}, we obtain
\begin{equation}
\int_{t_n}^{t_n+\Delta t} \left(\frac{dp}{dt}E_k-\frac{1}{2}\frac{dq}{dt}E_v\right)dt = \mathcal{E}(t_n+\Delta t)-\mathcal{E}(t_n).
\end{equation}
With this motivation, we define the discrete energy update:
\begin{equation}
    \mathcal{I}(\psi,\theta) := \int_{t_n}^{t_n+\theta\Delta t} \mathcal{J}(\psi,t, \theta) dt 
\end{equation}
where
\begin{equation}
    \mathcal{J}(\psi,t, \theta) :=
    \frac{dp}{dt}\bar{E}_k(\psi)-\frac{1}{2}\frac{dq}{dt}\bar{E}_v(\psi).
\end{equation}
This gives the expected change in the discrete energy over a time step of size $\theta \Delta t$.  Next, we use
the PR or MR method to ensure that the energy changes
by exactly this amount, taking
$\theta=1+\sum_i\gamma_i$ for MR and $\theta = \gamma$ for PR.
To apply multiple relaxation or projection relaxation, we replace equation \eqref{eq:MR-system_b} or eq.\eqref{eq:projRlx_opt} (respectively) with
\begin{equation}
\denergy(\psi^{n+1}_{\gamma}) - \denergy(\psi^n) =\mathcal{I}\left(\theta(\gamma)\right).
\end{equation}

We approximate the integral $\mathcal{I(\theta(\gamma))}$ above using the quadrature inherent to the Runge-Kutta method:
\begin{align}
    \mathcal{I(\theta(\gamma))} & \approx \theta \Delta t \sum_j b_j \mathcal{J}(\psi^{(j)},t_n+c_j \Delta t, \theta),
\end{align}
where $\mathcal{I}_j$ is the approximation of integrand at stage $j$ of the Runge-Kutta method.

To quantify the error in the energy balance, we monitor the following quantity:
\begin{equation}
\mathbf{E_{bal}}(t)= \mathcal{E}(t)-\mathcal{E}(t_{ini}) - \int_{t_{ini}}^{t} \left(\frac{dp}{dt}E_k-\frac{1}{2}\frac{dq}{dt}E_v\right)dt.
 \label{eq:EB}
\end{equation}

\subsection{Implementation}
Both MR and PR require the solution of algebraic equations
(to find $\gamma$) at each step.  After some testing, we have found that
Optimistix's \cite{optimistix2024} Newton solver\footnote{\url{https://docs.kidger.site/optimistix/api/root_find/}} is relatively fast and robust in this context.
In all cases, we use a tolerance of $10^{-14}$.
As discussed in the examples section, for MR the solver in rare cases fails to find a sufficiently accurate value of $\gamma$; in this case, we reduce the value of $\Delta t$ by half and redo the step, restoring $\Delta t$ to its original value at the next step.

\section{Numerical Studies}

\label{sec:numerical_cases}
We present numerical tests of the proposed methodology applied to three example problems.
We start with a 2D example that is conservative ($p$ and $q$ are constant), followed by a 2D non-conservative example ($p$ and $q$ vary in time), and finally a 3D non-conservative example.
For each example, we compare results obtained with the baseline method to those obtained with multiple relaxation (MR) and projection-relaxation (PR).

\subsection{2D Self-Gravitating Gaussians with Constant $p$ and $q$}
\label{subsec:two_gauss}
This example was used in Mocz et. al \cite{PhysRevD.97.083519} to study Schrödinger-Poisson–Vlasov-Poisson correspondence. The SP equations in this case are used to model a self-gravitating quantum superfluid and the coefficients $p$ and $q$ in eq. \eqref{eq:SP_a} are constants:
\begin{equation}
    p =  \frac{\hbar}{2m}   \qquad q = 4\pi G\frac{m}{\hbar}.
\end{equation}
The initial condition has density
\begin{equation}
\large
|\psi(t=0,x,y)|^2=A\left[\frac{1}{4}
+e^{-\frac{\left(x-\frac{5}{8}\right)^{2}+\left(y-\frac{1}{2}\right)^{2}}{2\sigma^{2}}}
+e^{-\frac{\left(x-\frac{3}{8}\right)^{2}+\left(y-\frac{1}{2}\right)^{2}}{2\sigma^{2}}}
\right],
\end{equation}
where $A=10^8 M_\odot Mpc^{-3}$, $\sigma=0.1$ Mpc, and $mc^2=8\times10^{-21}$eV, $G$ is the gravitational constant and $M_\odot$ stands for the solar mass. The boxsize is $L=1$ Mpc ($x\in[0,1]\times[0,1]$) and the final time is $T=1$ $Mpc(km/s)^{-1}$. We  discretize in space using $2048\times2048$ gridpoints. The initial phase of the wavefunction is zero (i.e. it is purely real). The system is evolved using ARK3(2)4L[2]SA \cite{kennedy2003additive}. 

\begin{figure}
    \centering
    \includegraphics[width=1.1\linewidth]{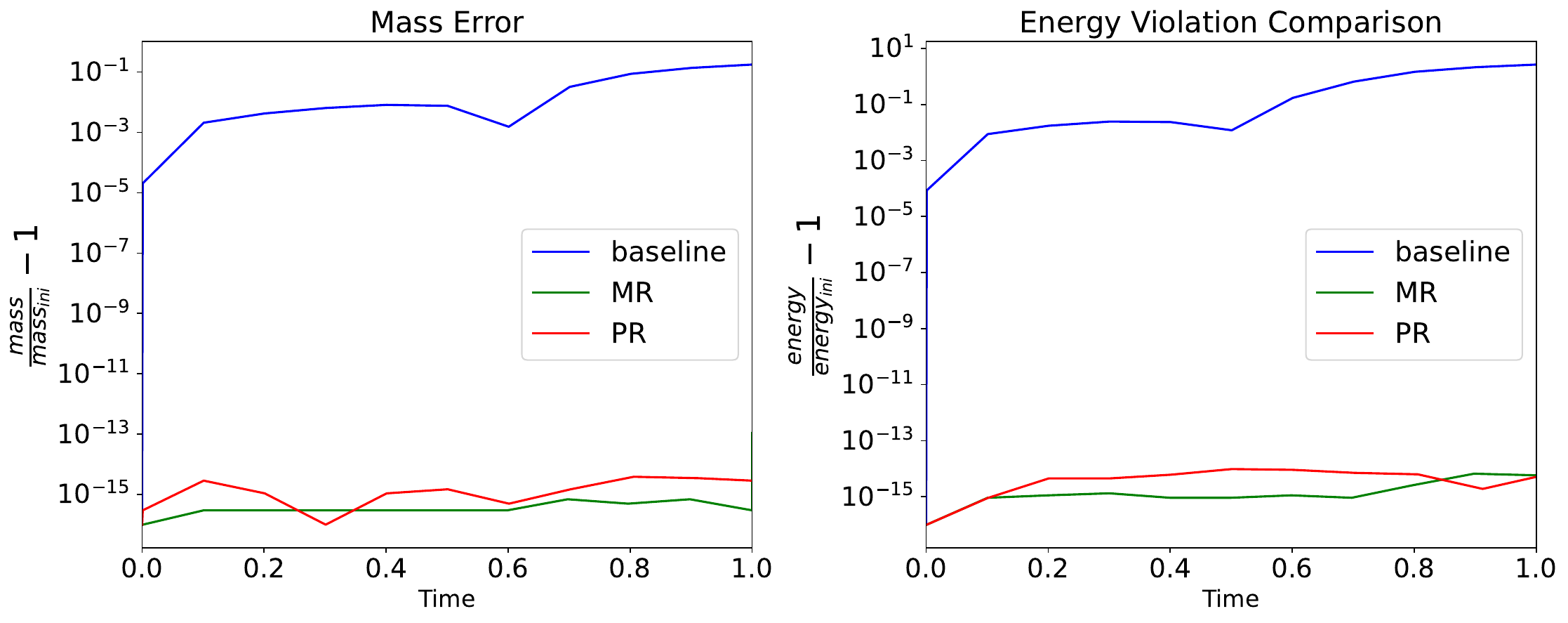}
    \caption{Mass change(left panel) and energy violation(right) evolution for 2D self-gravitating Gaussians \ref{subsec:two_gauss} with $\Delta t=10^{-3}$. Notice that the baseline method does not conserve mass or energy, while MR and PR do.}
    \label{fig:two_gauss_mass_energy}
\end{figure}

\begin{figure}
    \centering
    \includegraphics[width=1.0\linewidth]{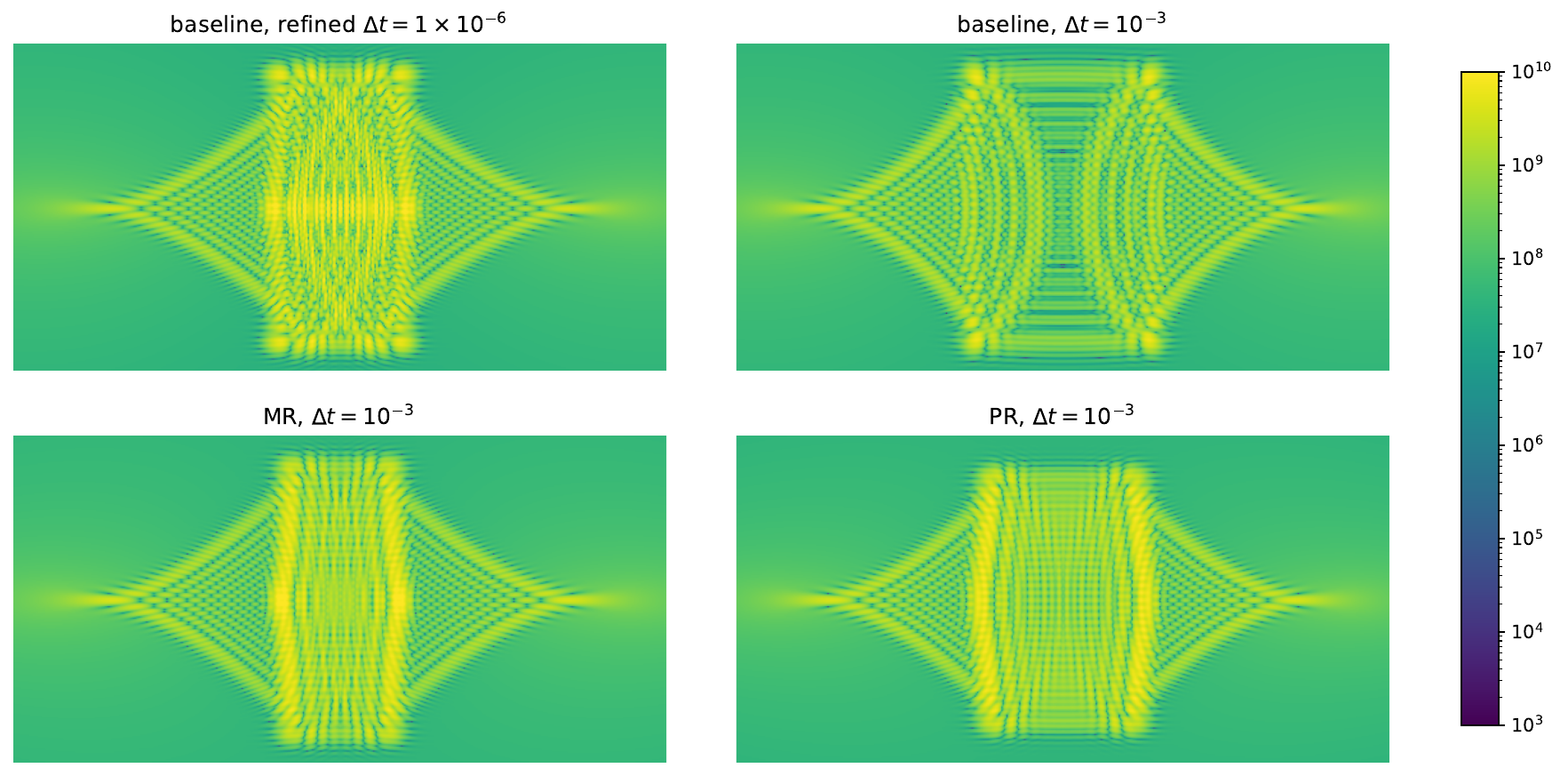}
    \caption{Density Plots for example \ref{subsec:two_gauss}: Top left panel shows final density plot for the reference case(refined $\Delta t=10^{-6}$). Here we see that non-relaxed case looks very different from any relaxed case and the solutions from two relaxation methods also differ from each other. But visually one can see that the relaxed cases are more in agreement with reference than non-relaxed one as the formation of a central core through interference  has started in both the relaxed cases as is the case with reference solution.}
    \label{fig:two_gauss_densityp001}
\end{figure}

\begin{figure}
    \centering
    \includegraphics[width=1.0\linewidth]{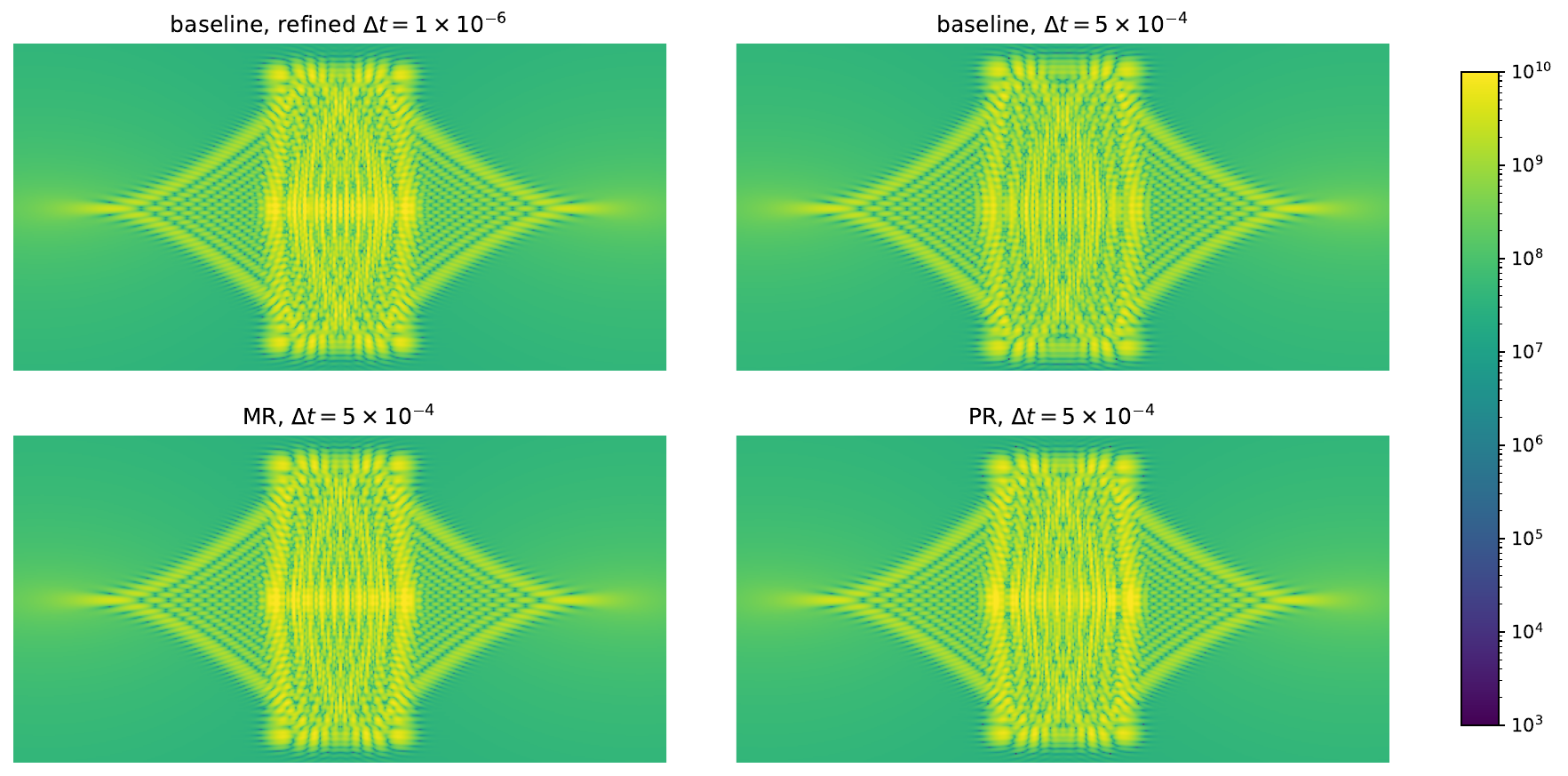}
    \caption{Same as previous plot \ref{fig:two_gauss_densityp001} but with refined timestep ($\Delta t=5\times10^{-4}$). In contrast to previous figure, here we see that all three cases start converging towards reference figure in top-left.  }
    \label{fig:two_gauss_densityp005}
\end{figure}

Since $p$ and $q$ are constant, the energy of the exact solution is conserved. However, the problem is highly nonlinear, with the density varying in space by 7 orders of magnitude. We show the evolution of the numerical mass and energy in Figure \ref{fig:two_gauss_mass_energy}. Relaxation conserves mass and energy (up to rounding errors) while the baseline method causes a relatively large change in mass and energy. Densities at the final time are plotted in Figures \ref{fig:two_gauss_densityp001} and \ref{fig:two_gauss_densityp005}. In these figures, we plot the solution only in the region ($[3/8,5/8]\times[7/16,9/16]$) in order to focus on the relevant structures that are formed. For Figure \ref{fig:two_gauss_densityp001} we use $\Delta t=10^{-3}$ and for Figure \ref{fig:two_gauss_densityp005} $\Delta t= 5\times 10^{-4}$. For comparison, we also show a more accurate solution computed with a finer time step of $10^{-6}$ in top-left panel of both the figures. This refined solution is visually in agreement with the panel 3 of Figure 4 from Mocz et. al \cite{PhysRevD.97.083519}.  In Figure \ref{fig:two_gauss_densityp001}, we see that the baseline method and the two relaxation methods give solutions that look qualitatively different from each other as well as from the reference solution. When we reduce the $\Delta t$ by half in the next figure (\ref{fig:two_gauss_densityp005}), we find that all solutions become more similar, suggesting that all 3 methods (baseline, MR, and PR) converge to the same solution as $\Delta t$ is refined. 

\begin{figure}
    \centering
    \includegraphics[width=1.0\linewidth]{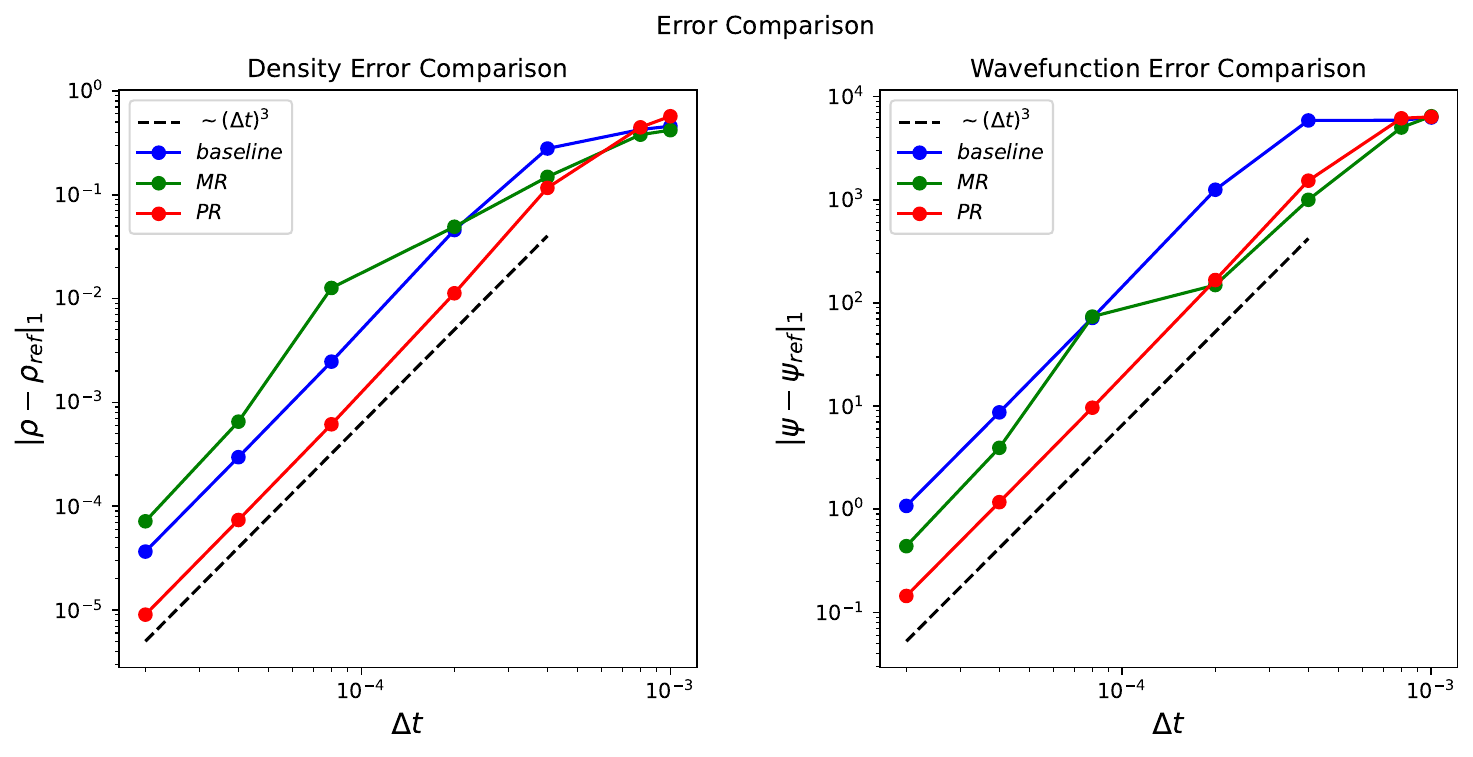}
    \caption{Error convergence for 2D example \ref{subsec:two_gauss}. A line with slope 3 is plotted in dashed black. The scheme used is 3rd order and both baseline and PR give 3rd order convergence. MR shows deviations from this slope.}
    \label{fig:two_gauss_err_conv}
\end{figure}

This is confirmed in Figure \ref{fig:two_gauss_err_conv}, where we plot errors as a function of $\Delta t$ for a range of step sizes.  Errors are computed with respect to
the solution obtained using $\Delta t = 10^{-6}$ with the baseline method. Projection-relaxation not only maintains the order of convergence of the baseline method (a 3rd order scheme in this case) but also reduces the errors compared to the baseline method. In comparison, multiple-relaxation gives errors in wavefunction comparable to the baseline method while the errors in density are slightly worse than the baseline. Also, multiple-relaxation requires solving a system of two equations, and in some cases the algebraic solver may fail to converge.  For projection-relaxation, which requires solving only a scalar equation, we have not encountered any cases of non-convergence.

In Table \ref{tab:ratio_comparison} we show the ratio of the run-time of the conservative (MR and PR) methods compared to the run-time of the non-conservative baseline method.
The cost of relaxation is not insignificant, since simply
evaluating the energy of a solution requires computing
Fourier transforms.  We see that these methods require
two to three times as much computational time as the baseline
method, with the PR method being significantly faster than MR.  Although this is a substantial cost, it is cheap
compared to the typical expense of using an implicit
time integrator.  We also show the fraction of time steps
for which the algebraic solver failed, indicating that failures may occur for MR when the timestep is relatively large.

\begin{table}[htbp]
\centering
\caption{Comparison of compute-time to the baseline and failure ratio for root-finding. 
The compute-time ratio is defined as 
$T_{\mathrm{method}}/T_{\mathrm{base}}$, 
and the failure ratio as 
$N_{\mathrm{fail}}/N_{\mathrm{steps}}$.}
\label{tab:ratio_comparison}
\begin{tabular}{|c|cc|cc|}
\hline
& \multicolumn{2}{c|}{Time ratio} & \multicolumn{2}{c|}{Failure ratio} \\
\cline{2-5}
$\Delta t$ & MR & PR & MR & PR \\
\hline
0.0001 & 2.5 & 1.9 & 0.054 & 0 \\
0.0004 & 2.4 & 1.9 & 0 & 0 \\
0.0008 & 2.6 & 2.1 & 0 & 0 \\
0.0010 & 2.7 & 2.1 & 0 & 0 \\
\hline
\end{tabular}
\end{table}

\subsection{2D Sine-Wave Collapse with Time-varying Coefficients } \label{subsec:sine_collapse}

\begin{figure}
    \centering
    \includegraphics[width=0.8\linewidth]{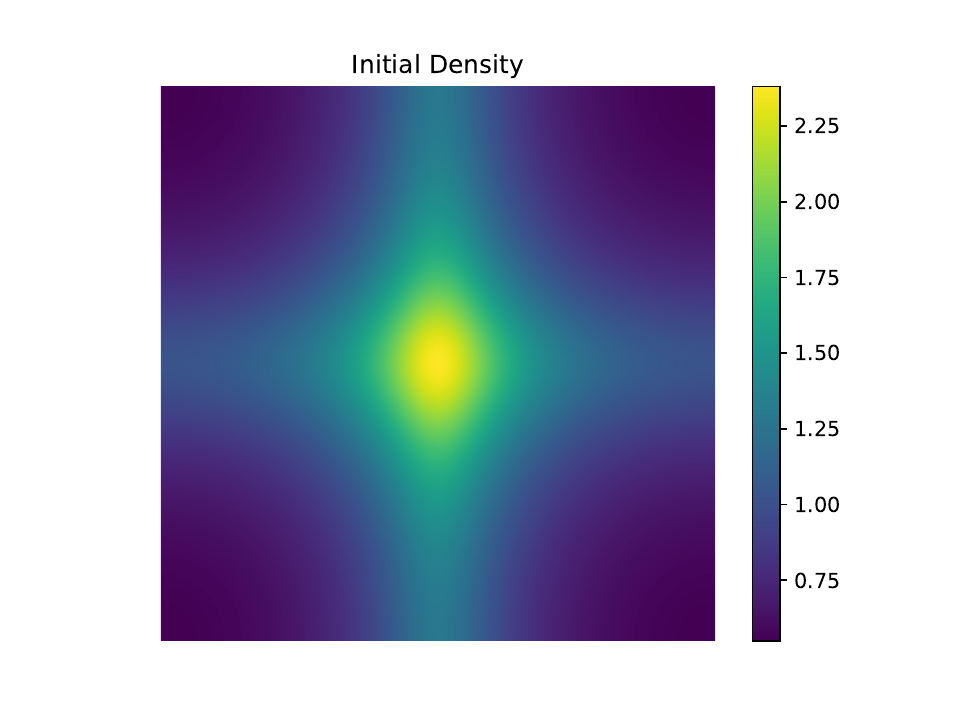}
    \caption{Initial density profile for 2D sine-wave collapse \ref{subsec:sine_collapse}}
    \label{fig:sine_collapse_ini}
\end{figure}

\begin{figure}
    \centering
    \includegraphics[width=1.1\linewidth]{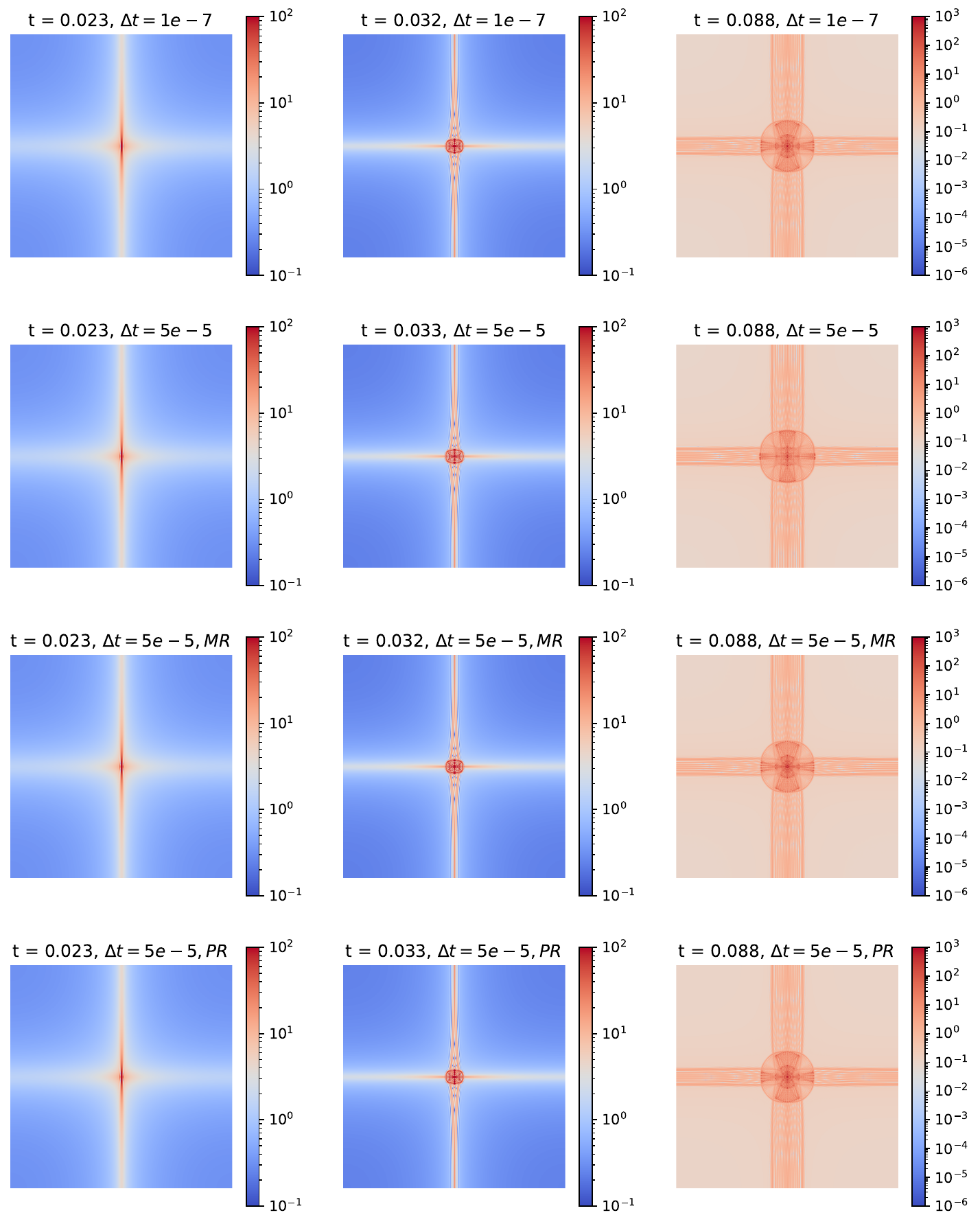}
    \caption{Sine-wave collapse at different times (changing across columns) and for different methods (changing across rows). Top row is for a refined $\Delta t=10^{-7}$ to be used as reference. The sine-waves collapse and interference patterns appear. By visual inspection, we can tell that relaxed cases give sharper features matching with the reference in comparison to the cases with the same $\Delta t$ but without relaxation.}
    \label{fig:sin_collapse}
\end{figure}

\begin{figure}
    \centering
    \includegraphics[width=1.\linewidth]{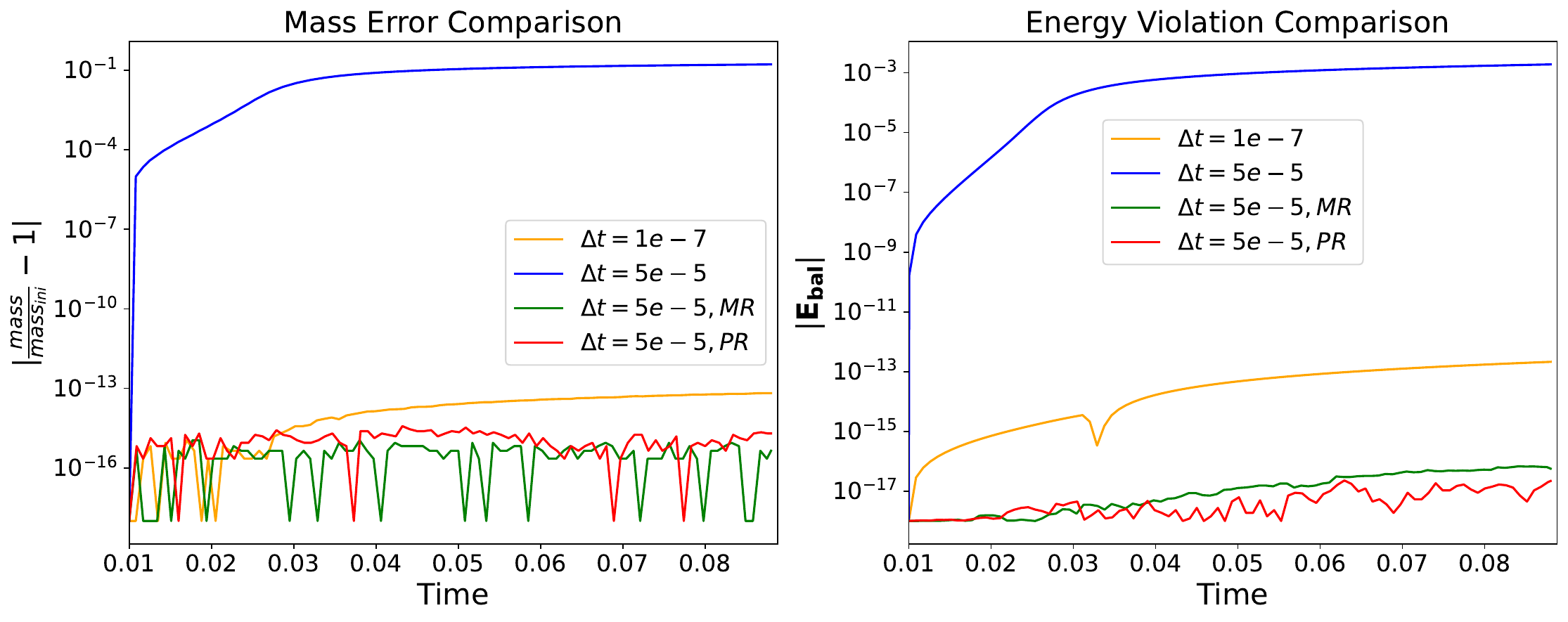}
    \caption{Mass (left panel) and energy-balance (right) for sine-wave collapse cases shown in Figure \ref{fig:sin_collapse}. By comparing curves for $\Delta t=5 \times 10^{-5}$ with and without relaxation, we can clearly see that relaxation upholds the invariance which is otherwise violated. $\Delta t=10^{-7}$ case also holds the conservation showing that the mass or energy violation for this spatial resolution ($2048\times2048$) is dominated by time discretization errors.}
    \label{fig:sine_collapse_mass_energy}
\end{figure}

This example comes from Athanassoulis et. al \cite{athanassoulis2023novel}.
As mentioned in the introduction, the SP equation can be used as an approximation to the Vlasov equations for a collisionless fluid, and hence it has been used as a technique for simulating dark matter dynamics. We use sine-wave collapse \cite{Sousbie_2016,Kopp_2017} as a demonstration test case to show the conservation properties and order of convergence of our proposed numerical methodology. This is a 2D example of a cosmological collapse where the initial condition is made of two sinusoidal perturbations along the two coordinate axes. The two perturbations have slightly different magnitudes. The process for initial condition generation has been described in \cite{Kopp_2017}. In order to do a consistent comparison, we borrow the initial conditions from authors of \cite{athanassoulis2023novel} and \cite{Kopp_2017}. The initial density is shown in Figure \ref{fig:sine_collapse_ini}.
Since this example is a toy model for collapse in expanding cosmology, the coefficients $p$ and $q$ in eq. \eqref{eq:SP_a} are time-dependent \cite{athanassoulis2023novel}:
\begin{equation}
    p(t) =  \frac{\epsilon}{2t^{3/2}}, \qquad q(t) = \frac{\beta}{\epsilon t^{1/2}}.
\end{equation}
Here $\beta = 3/2$ and $\epsilon=6\times10^{-5}$; note that the ``time" $t$ is actually the expansion scale factor, which is often denoted by $a$ in the literature.
We solve this case with ImEx scheme ARK4(3)6L[2]SA with and without relaxation. In Figure \ref{fig:sin_collapse}, we show the solutions obtained. 
We plot solutions at 3 different times, and each column shows the solution at a particular time. Each row shows solution computed with different methodology or parameters. Top row is a solution obtained with very small $\Delta t=10^{-7}$, that we use as reference solution. Second row  is  the baseline method with coarse $\Delta t=5\times10^{-5}$ and  and the last two rows show a conservative solutions obtained via the two relaxation methods with the same coarse time step.  While all four solutions exhibit very similar features, the relaxation solutions seem to more closely resemble the fine-grid solution at the final time, suggesting that relaxation improves the overall accuracy. In Figure \ref{fig:sine_collapse_mass_energy}, we show the evolution of mass and energy-balance for cases shown in Figure \ref{fig:sin_collapse}.

\begin{figure}
    \centering
    \includegraphics[width=1.1\linewidth]{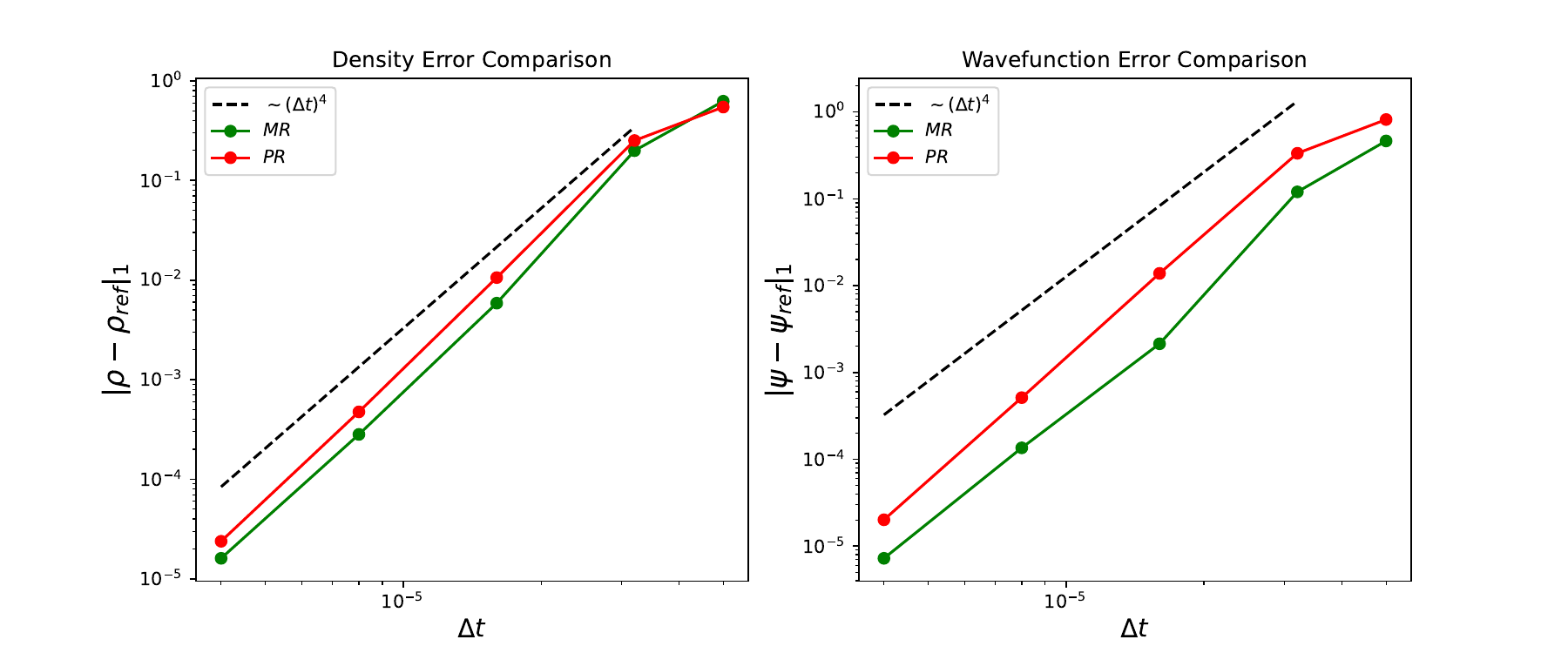}
    \caption{Self-convergence of solutions for a particular method as $\Delta t$ is changed. Error in density (left panel) and wavefunction (right) for 2D sine-wave case. Here we use as reference same type of method with very refined $\Delta t$ i.e. error of PR cases with different $\Delta t$s is calculated with reference to PR with very refined $\Delta t$. Similarly, smallest $\Delta t$ MR case is used for calculating errors for all MR cases in this figure. We see that convergence is around order $\sim 4$, as expected from the 4th order ImEx case used here.}
    \label{fig:sin_wave_order_convergence}
\end{figure}

\begin{figure}
    \centering
    \includegraphics[width=1.05\linewidth]{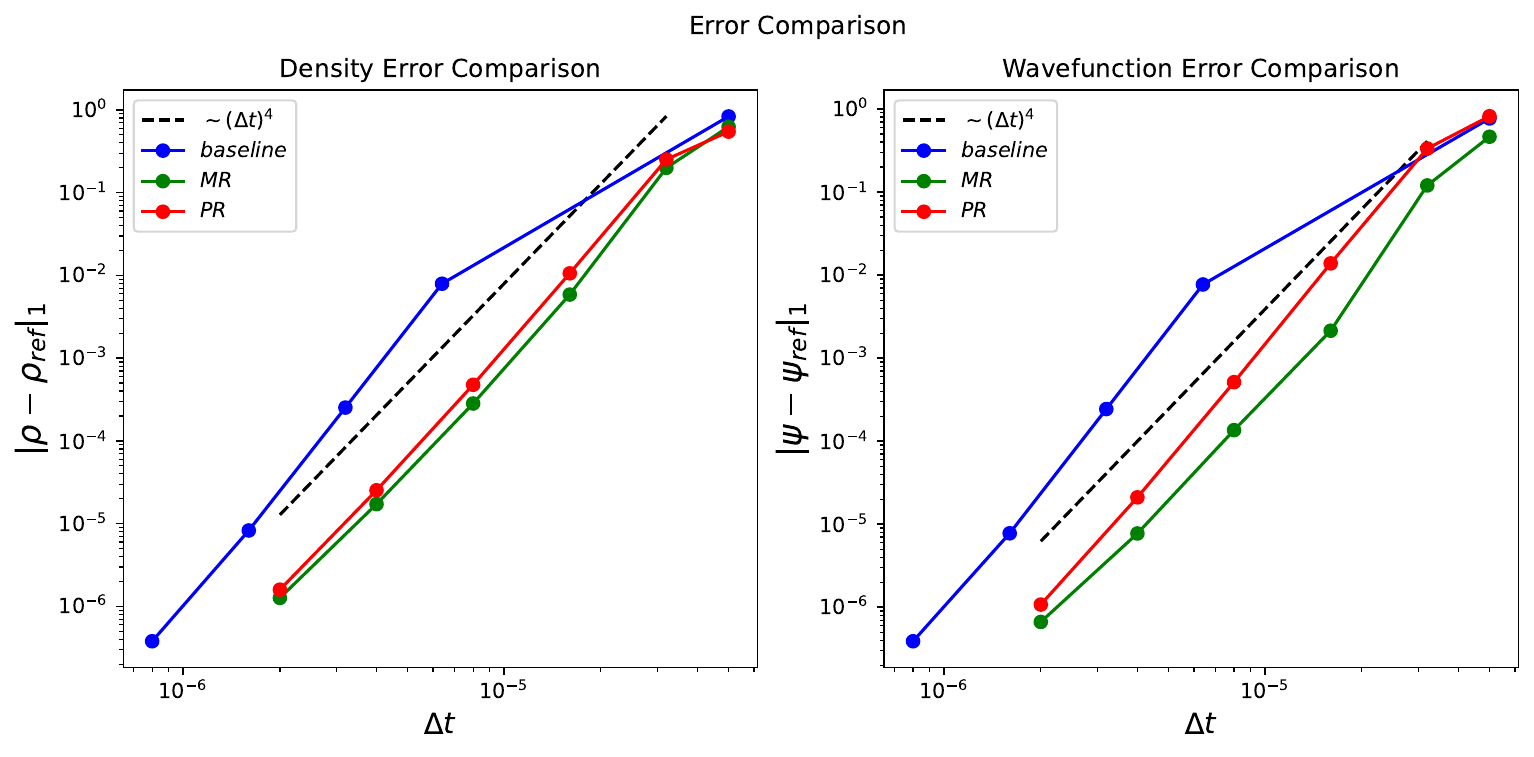}
    \caption{Convergence of solutions to same reference across the methods. In this figure we use a non-relaxed case with refined $\Delta t=10^{-7}$ as reference to calculate errors for all the points in this plot. The convergence is of order $\approx 4$ for both types of relaxation methods.}
    \label{fig:sine_wave_err}
\end{figure}

 Now we demonstrate that a 4th-order ImEx RK method maintains its order when combined with relaxation. We simulate this case with different timestep sizes($\Delta t$). Since we do not have an analytical expression for the solution in this case, we use the most refined simulation (smallest $\Delta t$) as a reference for checking the order of convergence. First we check the convergence of the relaxation method to its own refined $\Delta t$ simulation. For each relaxation, we take $\Delta t=10^{-7}$ combined with that type of relaxation  as reference solution for that relaxation method. We plot the errors in \ref{fig:sin_wave_order_convergence}. The ImEx scheme used is 4th order and we see from the dashed lines that the slope is around order $\sim 4$. All of them conserve mass and energy around order $10^{-14}$. Next, we take a refined $\Delta t=10^{-7}$ case without any relaxation as a reference solution for checking the error convergence. The results are shown in Figure \ref{fig:sine_wave_err}. This plot demonstrates two important points: (a) The methods with relaxation maintain their rough order of convergence. (b) The mass and energy conservation through relaxation improves the overall quality of solution as the curves for both relaxation methods are below the one without relaxation.

\subsection{3D Cosmological Simulation}
\label{subsec:cosmo_3d}
Here we present a 3D cosmological example taken from May et. al. \cite{10.1093/mnras/stab1764}.   Using the cosmological simulation software \verb|Gadget4| \cite{Springel_2021} we generate initial conditions (shown in the left panel of Figure \ref{fig:cosmo_3d_density_initial})  at time $t_{ini}=a=0.0078125$ (which corresponds to a redshift of $z=127$) and evolve until $a=1$ ($z=0$)\footnote{Note that redshift ($z$) is related to $a$ via $1+z=\frac{a_0}{a}$ where $a_0=1$ is the present value of the expansion factor $a$.}. The cosmological parameters are: $\Omega_m=0.3$, $\Omega_\Lambda=0.7$, and reduced Hubble parameter $h=0.7$.  
The spatial domain is a cube of width $L=1$ $Mpc/h$.  Although this is too small to be useful for detailed cosmological studies, we use it here as a demonstration that energy conservation via relaxation can enhance 3D astrophysical simulations. We use a grid of $256^3$ points in space.

\begin{figure}
    \centering
    \includegraphics[width=1\linewidth]{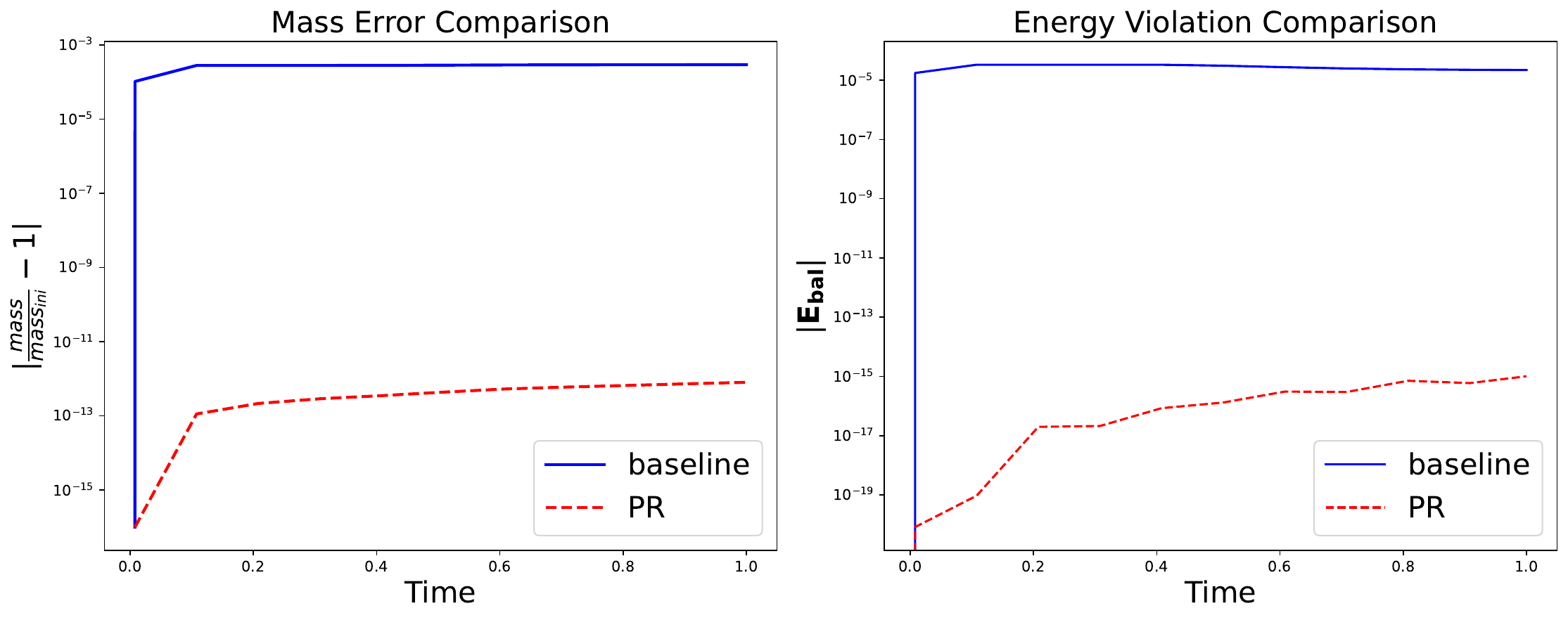}
    \caption{Mass change and energy violation for 3D cosmological example. While from previous figure \ref{fig:cosmo_3d_density}, we see that both relaxed and non-relaxed cases look almost indistinguishable, this figure shows there is huge difference for the conservation properties of the system.}
    \label{fig:cosmo_3d_mass_energy}
\end{figure}

\begin{figure}
    \centering
    \includegraphics[width=1.1\linewidth]{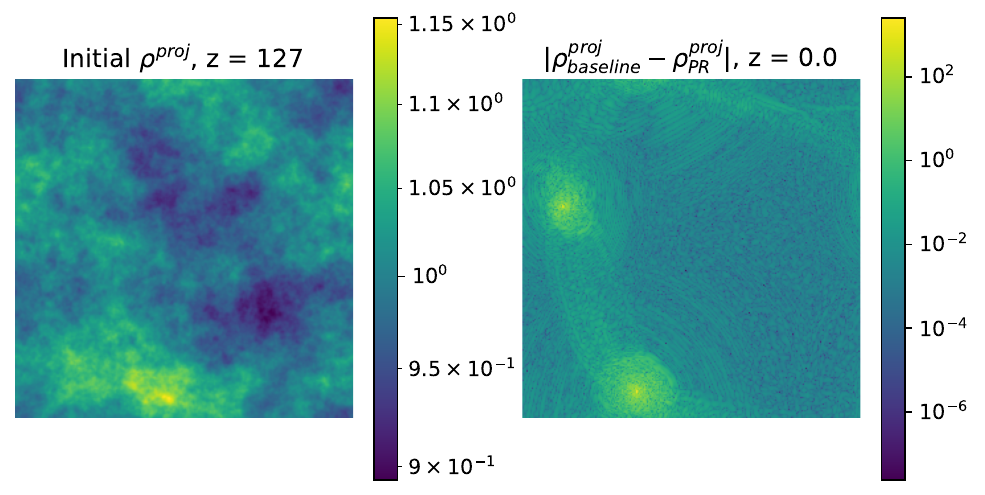}
    \caption{Projected initial density (left) and the difference between the projected densities calculated using two methods (right). The differences are higher in regions with higher densities. This indicates that effects of preserving mass and energy-balance may be more significant for halos (or nonlinear structures).}
    \label{fig:cosmo_3d_density_initial}
\end{figure}

\begin{figure}
    \centering
    \includegraphics[width=1.1\linewidth]{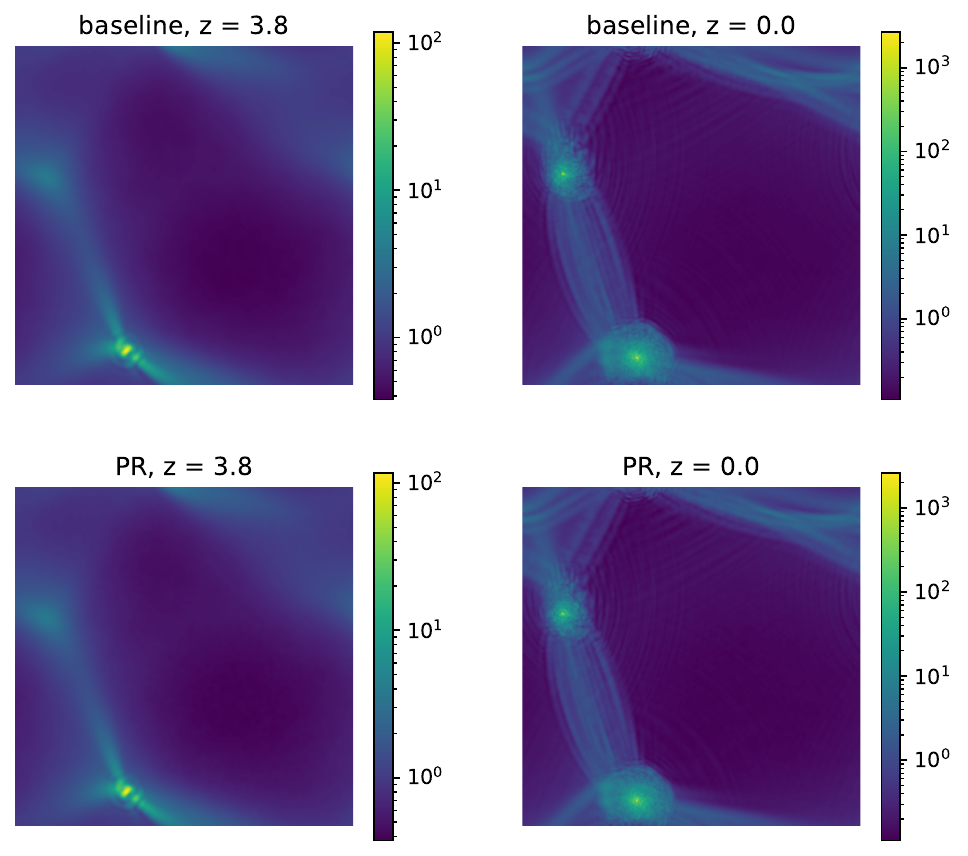}
    \caption{\textbf{3D cosmological example \ref{subsec:cosmo_expand}:}  Projected density ($\rho^{proj}$) for 2 different times (redshifts). The initial density ( Figure \ref{fig:cosmo_3d_density_initial}) is a gaussian random field distorted by linear perturbations. At later times, we see clustering with large-scale interference structure, as expected. At this scale, no differences between the methods are discernible.}
    \label{fig:cosmo_3d_density}
\end{figure}

As shown in Figure \ref{fig:cosmo_3d_mass_energy}, 
even with a time step of $\Delta t=5\times 10^{-5}$, the baseline method leads to mass and energy changes of about $10^{-4}$ and $10^{-5}$, respectively. In contrast, the PR method maintains
conservation to within rounding errors.
These differences are not obvious from a visual inspection
of the solutions, which are shown in
Figure \ref{fig:cosmo_3d_density} and exhibit the expected clumps and filament structures.  
The left panel of Figure \ref{fig:cosmo_3d_density_initial} shows initial (projected) density, while the right panel shows the difference between the projected densities at the final time ($z=0$) obtained by the baseline and PR methods. The differences are more pronounced near the high-density nonlinear structures (halos).

\begin{figure}
    \centering
    \includegraphics[width=1.\linewidth]{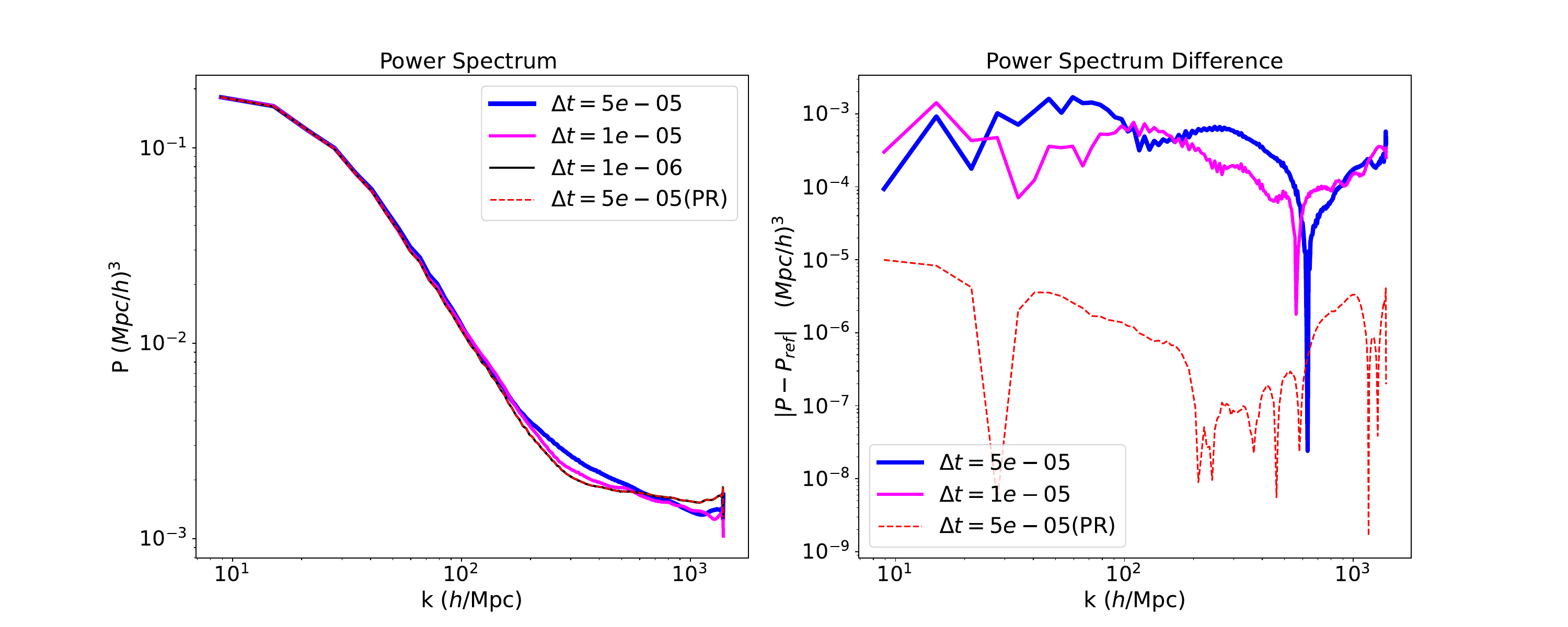}
    \caption{Power spectra for baseline method with different $\Delta t$  and PR ($\Delta t = 5\times 10^{-5}$, dashed line). Thicker lines correspond to larger time steps, $\Delta t$. The right panel shows the absolute difference from most refined baseline case $\Delta t=10^{-6}$.}
    \label{fig:cosmo_3d_power_diff}
\end{figure}

In the left panel of Figure \ref{fig:cosmo_3d_power_diff} we plot the power spectra obtained with the baseline method for three values of $\Delta t$, along with results using projection-relaxation with a step size corresponding to the largest one from the baseline experiments.  
In the right panel of the same figure, we take the most refined baseline case, $\Delta t = 10^{-6}$, as the reference and plot the differences between it and the power spectra of the other solutions, including the baseline method and the PR method with larger step sizes.
It is clear that the PR solution is more accurate, in the sense that it is closer to the reference solution by 1-3 orders of magnitude.

\section{Conclusion \& Prospects}
\label{sec:conclusion}

In this article, we have combined pseudospectral ImEx schemes with two types of relaxation methods to demonstrate a methodology that conserves mass and energy for Schr\"{o}dinger-Poisson systems. We showed that the method can be used to have higher-order in time numerical schemes while maintaining mass-energy conservation. It is also shown here that the semidescrete system obtained after pseudospectral space discretization using fourier transform, conserves mass and energy for SP equations. We have tested the methods on 2D and 3D test cases. The tests include both types of cases: (a) one where $p$ and $q$ are constants and energy is conserved and (b) cases with time-dependent $p(t)$ and $q(t)$ where energy satisfies a balance equation. We showed how to incorporate this energy balance in relaxation schemes. Between the two relaxation techniques considered here, we find that it is more easier for numerical solvers to solve the root-finding problem for projection-relaxation method than for multiple-relaxation methods. Hence, for 3D cases, we recommend using projection-relaxation methods.

Perhaps the most important contribution of this article is that it illustrates the usefulness and efficiency of a class of techniques that allow researchers to combine their favorite time evolution scheme with relaxation to get desired conservation properties. 
In the future, we plan to apply the methods developed here in two ways:
(a) studying the relaxation-enhanced kick-drift-kick type of operator-splitting methods that are popular with astrophysics/cosmology researchers; (b) conducting a thorough comparison of various mass- and energy-conserving schemes to study their effect on physically relevant features like halo-profile/soliton shape, statistical properties, etc. 

\section*{Acknowledgements}
Authors would like to thanks Prof. Theodoros Katsaounis for helping with the intial conditions for the sine-wave collapse example. Both authors were supported by funding from King Abdullah University of Science and Technology (KAUST). The computational work was done on the IBEX facility at KAUST.

\section*{Data Availability}
The code for the examples studied in this paper is available online at:\\
\url{https://github.com/manu0x/Conservative-Schrodinger-Poisson-ImEx}.

\clearpage

\bibliographystyle{plain}
\bibliography{ref}

@article{ringhofer2000discrete,
  title={Discrete {S}chr{\"o}dinger-{P}oisson systems preserving energy and mass},
  author={Ringhofer, Christian and Soler, Juan},
  journal={Applied Mathematics Letters},
  volume={13},
  number={7},
  pages={27--32},
  year={2000},
  publisher={Elsevier}
}

@article{ehrhardt2006fast,
  title={Fast calculation of energy and mass preserving solutions of {S}chr{\"o}dinger--{P}oisson systems on unbounded domains},
  author={Ehrhardt, Matthias and Zisowsky, Andrea},
  journal={Journal of computational and applied mathematics},
  volume={187},
  number={1},
  pages={1--28},
  year={2006},
  publisher={Elsevier}
}

@article{athanassoulis2023novel,
  title={A novel, structure-preserving, second-order-in-time relaxation scheme for {S}chr{\"o}dinger-{P}oisson systems},
  author={Athanassoulis, Agissilaos and Katsaounis, Theodoros and Kyza, Irene and Metcalfe, Stephen},
  journal={Journal of Computational Physics},
  volume={490},
  pages={112307},
  year={2023},
  publisher={Elsevier}
}

@article{nemati2025high,
  title={High-order numerical solution for solving multi-dimensional {S}chr{\"o}dinger-{P}oisson equation},
  author={Nemati, Maedeh and Abbaszadeh, Mostafa and Dehghan, Mehdi},
  journal={Applied Numerical Mathematics},
  year={2025},
  publisher={Elsevier}
}

@article{wang2025point,
  title={Point-wise error estimates of two mass- and energy-preserving schemes for two-dimensional {S}chr{\"o}dinger-{P}oisson equations},
  author={Wang, Jialing and Kong, Anxin and Wang, Tingchun and Cai, Wenjun},
  journal={Applied Numerical Mathematics},
  year={2025},
  publisher={Elsevier}
}

@article{Kopp_2017,
   title={Solving the {V}lasov equation in two spatial dimensions with the {S}chrödinger method},
   volume={96},
   ISSN={2470-0029},
   url={http://dx.doi.org/10.1103/PhysRevD.96.123532},
   DOI={10.1103/physrevd.96.123532},
   number={12},
   journal={Physical Review D},
   publisher={American Physical Society (APS)},
   author={Kopp, Michael and Vattis, Kyriakos and Skordis, Constantinos},
   year={2017},
   month=dec }

@ARTICLE{1993ApJ...416L..71W,
       author = {{Widrow}, Lawrence M. and {Kaiser}, Nick},
        title = {Using the {S}chrödinger Equation to Simulate Collisionless Matter},
      journal = {\apjl},
         year = 1993,
        month = oct,
       volume = {416},
        pages = {L71},
          doi = {10.1086/187073},
       adsurl = {https://ui.adsabs.harvard.edu/abs/1993ApJ...416L..71W}
}

@article{Sousbie_2016,
   title={Col{DICE}: A parallel {V}lasov–{P}oisson solver using moving adaptive simplicial tessellation},
   volume={321},
   ISSN={0021-9991},
   url={http://dx.doi.org/10.1016/j.jcp.2016.05.048},
   DOI={10.1016/j.jcp.2016.05.048},
   journal={Journal of Computational Physics},
   publisher={Elsevier BV},
   author={Sousbie, Thierry and Colombi, Stéphane},
   year={2016},
   month=sep, pages={644–697} }

@article{PhysRevD.97.083519,
  title = {{S}chr\"odinger-{P}oisson--{V}lasov-{P}oisson correspondence},
  author = {Mocz, Philip and Lancaster, Lachlan and Fialkov, Anastasia and Becerra, Fernando and Chavanis, Pierre-Henri},
  journal = {Phys. Rev. D},
  volume = {97},
  issue = {8},
  pages = {083519},
  numpages = {17},
  year = {2018},
  month = {Apr},
  publisher = {American Physical Society},
  doi = {10.1103/PhysRevD.97.083519},
  url = {https://link.aps.org/doi/10.1103/PhysRevD.97.083519}
}

@article{MARSH20161,
title = {Axion Cosmology},
journal = {Physics Reports},
volume = {643},
pages = {1-79},
year = {2016},
note = {Axion Cosmology},
issn = {0370-1573},
doi = {https://doi.org/10.1016/j.physrep.2016.06.005},
url = {https://www.sciencedirect.com/science/article/pii/S0370157316301557},
author = {David J.E. Marsh}
}

@article{PhysRevLett.85.1158,
  title = {Fuzzy Cold Dark Matter: The Wave Properties of Ultralight Particles},
  author = {Hu, Wayne and Barkana, Rennan and Gruzinov, Andrei},
  journal = {Phys. Rev. Lett.},
  volume = {85},
  issue = {6},
  pages = {1158--1161},
  numpages = {0},
  year = {2000},
  month = {Aug},
  publisher = {American Physical Society},
  doi = {10.1103/PhysRevLett.85.1158},
  url = {https://link.aps.org/doi/10.1103/PhysRevLett.85.1158}
}

@article{Zhang_2019,
   title={Cosmological Simulation for Fuzzy Dark Matter Model},
   volume={5},
   ISSN={2296-987X},
   url={http://dx.doi.org/10.3389/fspas.2018.00048},
   DOI={10.3389/fspas.2018.00048},
   journal={Frontiers in Astronomy and Space Sciences},
   publisher={Frontiers Media SA},
   author={Zhang, Jiajun and Liu, Hantao and Chu, Ming-Chung},
   year={2019},
   month=jan }

@article{PhysRevLett.113.261302,
  title = {Understanding the Core-Halo Relation of Quantum Wave Dark Matter from 3D Simulations},
  author = {Schive, Hsi-Yu and Liao, Ming-Hsuan and Woo, Tak-Pong and Wong, Shing-Kwong and Chiueh, Tzihong and Broadhurst, Tom and Hwang, W-Y. Pauchy},
  journal = {Phys. Rev. Lett.},
  volume = {113},
  issue = {26},
  pages = {261302},
  numpages = {6},
  year = {2014},
  month = {Dec},
  publisher = {American Physical Society},
  doi = {10.1103/PhysRevLett.113.261302},
  url = {https://link.aps.org/doi/10.1103/PhysRevLett.113.261302}
}

@article{Mocz_2017,
   title={Galaxy formation with {BECDM} – {I}. {T}urbulence and relaxation of idealized haloes},
   volume={471},
   ISSN={1365-2966},
   url={http://dx.doi.org/10.1093/mnras/stx1887},
   DOI={10.1093/mnras/stx1887},
   number={4},
   journal={Monthly Notices of the Royal Astronomical Society},
   publisher={Oxford University Press (OUP)},
   author={Mocz, Philip and Vogelsberger, Mark and Robles, Victor H. and Zavala, Jesús and Boylan-Kolchin, Michael and Fialkov, Anastasia and Hernquist, Lars},
   year={2017},
   month=jul, pages={4559–4570} }

@article{Edwards_2018,
   title={Py{U}ltra{L}ight: a pseudo-spectral solver for ultralight dark matter dynamics},
   volume={2018},
   ISSN={1475-7516},
   url={http://dx.doi.org/10.1088/1475-7516/2018/10/027},
   DOI={10.1088/1475-7516/2018/10/027},
   number={10},
   journal={Journal of Cosmology and Astroparticle Physics},
   publisher={IOP Publishing},
   author={Edwards, Faber and Kendall, Emily and Hotchkiss, Shaun and Easther, Richard},
   year={2018},
   month=oct, pages={027–027} }

@article{Schive_2014,
   title={Cosmic structure as the quantum interference of a coherent dark wave},
   volume={10},
   ISSN={1745-2481},
   url={http://dx.doi.org/10.1038/nphys2996},
   DOI={10.1038/nphys2996},
   number={7},
   journal={Nature Physics},
   publisher={Springer Science and Business Media LLC},
   author={Schive, Hsi-Yu and Chiueh, Tzihong and Broadhurst, Tom},
   year={2014},
   month=jun, pages={496–499} }

@article{de_Blok_2009,
   title={The Core‐Cusp Problem},
   volume={2010},
   ISSN={1687-7977},
   url={http://dx.doi.org/10.1155/2010/789293},
   DOI={10.1155/2010/789293},
   number={1},
   journal={Advances in Astronomy},
   publisher={Wiley},
   author={de Blok, W. J. G.},
   editor={Brinks, Elias},
   year={2009},
   month=nov }

@article{10.1093/mnras/stv1050,
    author = {Marsh, David J. E. and Pop, Ana-Roxana},
    title = {Axion dark matter, solitons and the cusp–core problem},
    journal = {Monthly Notices of the Royal Astronomical Society},
    volume = {451},
    number = {3},
    pages = {2479-2492},
    year = {2015},
    month = {06},
    issn = {0035-8711},
    doi = {10.1093/mnras/stv1050},
    url = {https://doi.org/10.1093/mnras/stv1050},
    eprint = {https://academic.oup.com/mnras/article-pdf/451/3/2479/4001975/stv1050.pdf},
}

@article{Woo_2009,
doi = {10.1088/0004-637X/697/1/850},
url = {https://doi.org/10.1088/0004-637X/697/1/850},
year = {2009},
month = {may},
publisher = {The American Astronomical Society},
volume = {697},
number = {1},
pages = {850},
author = {Woo, Tak-Pong and Chiueh, Tzihong},
title = {High-resolution Simulation on Structure Formation with Extremely Light Bosonic Dark Matter},
journal = {The Astrophysical Journal}
}

@misc{schive2025fuzzydarkmattersimulations,
      title={Fuzzy dark matter simulations}, 
      author={Hsi-Yu Schive},
      year={2025},
      eprint={2509.23231},
      archivePrefix={arXiv},
      primaryClass={astro-ph.CO},
      url={https://arxiv.org/abs/2509.23231}, 
}

@article{PhysRevLett.121.151301,
  title = {Gravitational {B}ose-{E}instein Condensation in the Kinetic Regime},
  author = {Levkov, D. G. and Panin, A. G. and Tkachev, I. I.},
  journal = {Phys. Rev. Lett.},
  volume = {121},
  issue = {15},
  pages = {151301},
  numpages = {5},
  year = {2018},
  month = {Oct},
  publisher = {American Physical Society},
  doi = {10.1103/PhysRevLett.121.151301},
  url = {https://link.aps.org/doi/10.1103/PhysRevLett.121.151301}
}

@article{PhysRevD.102.083518,
  title = {Simulating mixed fuzzy and cold dark matter},
  author = {Schwabe, Bodo and Gosenca, Mateja and Behrens, Christoph and Niemeyer, Jens C. and Easther, Richard},
  journal = {Phys. Rev. D},
  volume = {102},
  issue = {8},
  pages = {083518},
  numpages = {10},
  year = {2020},
  month = {Oct},
  publisher = {American Physical Society},
  doi = {10.1103/PhysRevD.102.083518},
  url = {https://link.aps.org/doi/10.1103/PhysRevD.102.083518}
}

@ARTICLE{2020A&A...641A.107M,
       author = {{Mina}, Mattia and {Mota}, David F. and {Winther}, Hans A.},
        title = {{SCALAR}: an {AMR} code to simulate axion-like dark matter models},
      journal = {\aap},
         year = 2020,
        month = sep,
       volume = {641},
          eid = {A107},
        pages = {A107},
          doi = {10.1051/0004-6361/201936272},
archivePrefix = {arXiv},
       eprint = {1906.12160},
 primaryClass = {physics.comp-ph},
       adsurl = {https://ui.adsabs.harvard.edu/abs/2020A&A...641A.107M}
}

@article{10.1093/mnras/stab1764,
    author = {May, Simon and Springel, Volker},
    title = {Structure formation in large-volume cosmological simulations of fuzzy dark matter: impact of the non-linear dynamics},
    journal = {Monthly Notices of the Royal Astronomical Society},
    volume = {506},
    number = {2},
    pages = {2603-2618},
    year = {2021},
    month = {06},
    issn = {0035-8711},
    doi = {10.1093/mnras/stab1764},
    url = {https://doi.org/10.1093/mnras/stab1764},
    eprint = {https://academic.oup.com/mnras/article-pdf/506/2/2603/39207201/stab1764.pdf},
}

@PHDTHESIS{1961PhDT.........2I,
       author = {{Irvine}, William Michael},
        title = {Local Irregularities in a Universe Satisfying the Cosmological Principle.},
       school = {Harvard University, Massachusetts},
         year = 1961,
        month = jan,
       adsurl = {https://ui.adsabs.harvard.edu/abs/1961PhDT.........2I}
}

@ARTICLE{1963ApJ...138..174L,
       author = {{Layzer}, David},
        title = {A Preface to Cosmogony. {I}. The Energy Equation and the Virial Theorem for Cosmic Distributions.},
      journal = {\apj},
         year = 1963,
        month = jul,
       volume = {138},
        pages = {174},
          doi = {10.1086/147625},
       adsurl = {https://ui.adsabs.harvard.edu/abs/1963ApJ...138..174L}
}

@article{PAREDES2020132301,
title = {From optics to dark matter: A review on nonlinear {S}chrödinger–{P}oisson systems},
journal = {Physica D: Nonlinear Phenomena},
volume = {403},
pages = {132301},
year = {2020},
issn = {0167-2789},
doi = {https://doi.org/10.1016/j.physd.2019.132301},
url = {https://www.sciencedirect.com/science/article/pii/S0167278919307079},
author = {Angel Paredes and David N. Olivieri and Humberto Michinel}
}

@article{PhysRevD.92.124008,
  title = {Stability of self-gravitating {B}ose-{E}instein condensates},
  author = {Schroven, Kris and List, Meike and L\"ammerzahl, Claus},
  journal = {Phys. Rev. D},
  volume = {92},
  issue = {12},
  pages = {124008},
  numpages = {13},
  year = {2015},
  month = {Dec},
  publisher = {American Physical Society},
  doi = {10.1103/PhysRevD.92.124008},
  url = {https://link.aps.org/doi/10.1103/PhysRevD.92.124008}
}

@article{PhysRev.187.1767,
  title = {Systems of Self-Gravitating Particles in {G}eneral {R}elativity and the Concept of an Equation of State},
  author = {Ruffini, Remo and Bonazzola, Silvano},
  journal = {Phys. Rev.},
  volume = {187},
  issue = {5},
  pages = {1767--1783},
  numpages = {0},
  year = {1969},
  month = {Nov},
  publisher = {American Physical Society},
  doi = {10.1103/PhysRev.187.1767},
  url = {https://link.aps.org/doi/10.1103/PhysRev.187.1767}
}

@book{10.5555/89817, author = {Markowich, Peter A. and Ringhofer, Christian A. and Schmeiser, Christian}, title = {Semiconductor equations}, year = {1990}, isbn = {3211821570}, publisher = {Springer-Verlag}, address = {Berlin, Heidelberg} }

@article{Ranocha_2020,
   title={General relaxation methods for initial-value problems with application to multistep schemes},
   volume={146},
   ISSN={0945-3245},
   url={http://dx.doi.org/10.1007/s00211-020-01158-4},
   DOI={10.1007/s00211-020-01158-4},
   number={4},
   journal={Numerische Mathematik},
   publisher={Springer Science and Business Media LLC},
   author={Ranocha, Hendrik and Lóczi, Lajos and Ketcheson, David I.},
   year={2020},
   month=oct, pages={875–906} }

@misc{ranocha2025highordermassenergyconservingmethods,
      title="{High-order mass- and energy-conserving methods for the nonlinear {S}chr\"odinger equation and its hyperbolization}", 
      author={Hendrik Ranocha and David I. Ketcheson},
      year={2025},
      eprint={2510.14335},
      archivePrefix={arXiv},
      primaryClass={math.NA},
      url={https://arxiv.org/abs/2510.14335}, 
}

@article{doi:10.1137/23M1598118,
author = {Biswas, Abhijit and Ketcheson, David I.},
title = {Accurate Solution of the Nonlinear {S}chrödinger Equation via Conservative Multiple-Relaxation {I}m{E}x Methods},
journal = {SIAM Journal on Scientific Computing},
volume = {46},
number = {6},
pages = {A3827-A3848},
year = {2024},
doi = {10.1137/23M1598118},

URL = { 
    
        https://doi.org/10.1137/23M1598118
    
    

},
eprint = { 
    
        https://doi.org/10.1137/23M1598118
    
    

}
}

@article{SANZSERNA1982199,
title = {An explicit finite-difference scheme with exact conservation properties},
journal = {Journal of Computational Physics},
volume = {47},
number = {2},
pages = {199-210},
year = {1982},
issn = {0021-9991},
doi = {https://doi.org/10.1016/0021-9991(82)90074-2},
url = {https://www.sciencedirect.com/science/article/pii/0021999182900742},
author = {J.M Sanz-Serna}
}

@article{SANZSERNA1983273,
title = {A method for the integration in time of certain partial differential equations},
journal = {Journal of Computational Physics},
volume = {52},
number = {2},
pages = {273-289},
year = {1983},
issn = {0021-9991},
doi = {https://doi.org/10.1016/0021-9991(83)90031-1},
url = {https://www.sciencedirect.com/science/article/pii/0021999183900311},
author = {J.M Sanz-Serna and V.S Manoranjan}
}

@misc{ketcheson2019relaxationrungekuttamethodsconservation,
      title={Relaxation {R}unge-{K}utta Methods: Conservation and stability for Inner-Product Norms}, 
      author={David I. Ketcheson},
      year={2019},
      eprint={1905.09847},
      archivePrefix={arXiv},
      primaryClass={math.NA},
      url={https://arxiv.org/abs/1905.09847}, 
}

@article{https://doi.org/10.1111/sapm.70129,
author = {Biswas, Abhijit and Busaleh, Laila S. and Ketcheson, David I. and Muñoz-Moncayo, Carlos and Rajvanshi, Manvendra},
title = {A Hyperbolic Approximation of the Nonlinear {S}chrödinger Equation},
journal = {Studies in Applied Mathematics},
volume = {155},
number = {4},
pages = {e70129},
doi = {https://doi.org/10.1111/sapm.70129},
url = {https://onlinelibrary.wiley.com/doi/abs/10.1111/sapm.70129},
eprint = {https://onlinelibrary.wiley.com/doi/pdf/10.1111/sapm.70129},
year = {2025}
}

@article{Springel_2021,
   title={Simulating cosmic structure formation with the {GADGET-4} code},
   volume={506},
   ISSN={1365-2966},
   url={http://dx.doi.org/10.1093/mnras/stab1855},
   DOI={10.1093/mnras/stab1855},
   number={2},
   journal={Monthly Notices of the Royal Astronomical Society},
   publisher={Oxford University Press (OUP)},
   author={Springel, Volker and Pakmor, Rüdiger and Zier, Oliver and Reinecke, Martin},
   year={2021},
   month=jul, pages={2871–2949} }

@article{kennedy2003additive,
	author = {Kennedy, C. A. and Carpenter, M. H.},
	journal = {Applied Numerical Mathematics},
	pages = {139--181},
	title = {Additive {R}unge-{K}utta Schemes for Convection-Diffusion-Reaction Equations},
	volume = {44},
	year = {2003}
}

@book{doi:10.1137/1.9781611978209,
author = {Boscarino, Sebastiano and Pareschi, Lorenzo and Russo, Giovanni},
title = {Implicit-{E}xplicit Methods for Evolutionary Partial Differential Equations},
publisher = {Society for Industrial and Applied Mathematics},
year = {2024},
doi = {10.1137/1.9781611978209},
address = {Philadelphia, PA},
edition   = {},
URL = {https://epubs.siam.org/doi/abs/10.1137/1.9781611978209},
eprint = {https://epubs.siam.org/doi/pdf/10.1137/1.9781611978209}
}

@article{10.1093/imanum/drz067,
    author = {Besse, Christophe and Descombes, Stéphane and Dujardin, Guillaume and Lacroix-Violet, Ingrid},
    title = {Energy-preserving methods for nonlinear {S}chrödinger equations},
    journal = {IMA Journal of Numerical Analysis},
    volume = {41},
    number = {1},
    pages = {618-653},
    year = {2020},
    month = {06},
    issn = {0272-4979},
    doi = {10.1093/imanum/drz067},
    url = {https://doi.org/10.1093/imanum/drz067},
    eprint = {https://academic.oup.com/imajna/article-pdf/41/1/618/35971306/drz067.pdf},
}

@Article{SAV,
title = {{SAV} {G}alerkin-{L}egendre spectral method for the nonlinear {S}chrödinger-{P}oisson equations},
journal = {Electronic Research Archive},
volume = {30},
number = {3},
pages = {943-960},
year = {2022},
issn = {2688-1594},
doi = {10.3934/era.2022049},
url = {https://www.aimspress.com/article/doi/10.3934/era.2022049},
author = {Chunye Gong and Mianfu She and Wanqiu Yuan and Dan Zhao},
}

@ARTICLE{8466572,
  author={Cheng, Xiujun and Chen, Xiaoli and Li, Dongfang},
  journal={IEEE Access}, 
  title={Effective Mass and Energy Recovery by Conserved Compact Finite Difference Schemes}, 
  year={2018},
  volume={6},
  number={},
  pages={52336-52344},
  doi={10.1109/ACCESS.2018.2870254}}

@article{husimi,
  title={Some Formal Properties of the Density Matrix},
  author={K. Husimi},
  journal={Proceedings of the Physico-Mathematical Society of Japan. 3rd Series},
  volume={22},
  number={4},
  pages={264-314},
  year={1940},
  doi={10.11429/ppmsj1919.22.4_264}
}

@article{PhysRevD.84.043531,
  title = {Mass-radius relation of {N}ewtonian self-gravitating {B}ose-{E}instein condensates with short-range interactions. I. {A}nalytical results},
  author = {Chavanis, Pierre-Henri},
  journal = {Phys. Rev. D},
  volume = {84},
  issue = {4},
  pages = {043531},
  numpages = {27},
  year = {2011},
  month = {Aug},
  publisher = {American Physical Society},
  doi = {10.1103/PhysRevD.84.043531},
  url = {https://link.aps.org/doi/10.1103/PhysRevD.84.043531}
}

@article{optimistix2024,
    title={Optimistix: modular optimisation in {JAX} and {E}quinox},
    author={Jason Rader and Terry Lyons and Patrick Kidger},
    journal={arXiv:2402.09983},
    year={2024},
    url={https://docs.kidger.site/optimistix/api/root_find/},
}

@article{M2AN_2017__51_4_1245_0,
     author = {Auzinger, Winfried and Kassebacher, Thomas and Koch, Othmar and Thalhammer, Mechthild},
     title = {Convergence of a {Strang} splitting finite element discretization for the {Schr\"odinger{\textendash}Poisson} equation},
     journal = {ESAIM: Mathematical Modelling and Numerical Analysis },
     pages = {1245--1278},
     year = {2017},
     publisher = {EDP-Sciences},
     volume = {51},
     number = {4},
     doi = {10.1051/m2an/2016059},
     zbl = {1379.65071},
     mrnumber = {3702412},
     language = {en},
     url = {https://www.numdam.org/articles/10.1051/m2an/2016059/}
}

@ARTICLE{Bao2003809,
	author = {Bao, Weizhu and Mauser, N.J. and Stimming, H.P.},
	title = {Effective one particle quantum dynamics of electrons:: A numerical study of the {Schr\"odinger-Poisson}-{X$\alpha$} model},
	year = {2003},
	journal = {Communications in Mathematical Sciences},
	volume = {1},
	number = {4},
	pages = {809 – 828},
	doi = {10.4310/CMS.2003.v1.n4.a8},
	url = {https://www.scopus.com/inward/record.uri?eid=2-s2.0-11244292274&doi=10.4310%2fCMS.2003.v1.n4.a8&partnerID=40&md5=f0d844d8eb05f479de2e24c7be6c46d8},
	type = {Article},
	publication_stage = {Final},
	source = {Scopus},
	note = {Cited by: 54; All Open Access, Bronze Open Access}
}

@ARTICLE{Cheng2022523,
	author = {Cheng, Ronghua and Wu, Liping and Pang, Chunping and Wang, Hanquan},
	title = {A {F}ourier collocation method for {Schr\"odinger-Poisson} system with perfectly matched layer},
	year = {2022},
	journal = {Communications in Mathematical Sciences},
	volume = {20},
	number = {2},
	pages = {523 – 542},
	doi = {10.4310/CMS.2022.v20.n2.a10},
	url = {https://www.scopus.com/inward/record.uri?eid=2-s2.0-85124152539&doi=10.4310%2fCMS.2022.v20.n2.a10&partnerID=40&md5=43e592de8ca62cd0c9bcd8afe810b74d},
	type = {Article},
	publication_stage = {Final},
	source = {Scopus},
	note = {Cited by: 6}
}
\end{document}